\documentclass[%
aip,
amsmath,amssymb,
preprint,%
]{revtex4-2}

\usepackage{graphicx}% Include figure files
\usepackage{dcolumn}% Align table columns on decimal point
\usepackage{bm}% bold math
\usepackage{physics}
\usepackage{color}
\usepackage[utf8]{inputenc}
\usepackage[T1]{fontenc}
\usepackage{mathptmx}
\usepackage{setspace}
\usepackage[margin=1in]{geometry}
\usepackage{physics}
\usepackage{color}
\usepackage{multirow}
\usepackage{float}
\usepackage{adjustbox}
\usepackage{booktabs, tabularx} % added_AS for table in text
\newcolumntype{C}{>{\centering\arraybackslash}X}% added_AS for table in text

\newcommand\scalemath[2]{\scalebox{#1}{\mbox{\ensuremath{\displaystyle #2}}}}

\begin{document}

\title {Vibronic Resonance Along Effective Modes Mediates Selective Energy Transfer in Excitonically Coupled Aggregates}

\author{Sanjoy Patra}
\affiliation{ 
	Solid State and Structural Chemistry Unit, Indian Institute of Science, Bangalore, Karnataka 560012, India
}%
\author{Vivek Tiwari}%
\email{Author to whom correspondence should be addressed:vivektiwari@iisc.ac.in}
\affiliation{ 
	Solid State and Structural Chemistry Unit, Indian Institute of Science, Bangalore, Karnataka 560012, India
}%

\date{\today}% It is always \today, today,

%\doublespacing

\begin{abstract}
We recently proposed effective normal modes for excitonically coupled aggregates which exactly transform the energy transfer Hamiltonian into a sum of one-dimensional Hamiltonians along the effective normal modes. Identifying physically meaningful vibrational motions which maximally promote vibronic mixing suggested an interesting possibility of leveraging vibrational-electronic resonance for mediating selective energy transfer. Here we expand on the effective mode approach elucidating its iterative nature for successively larger aggregates, and extend the idea of mediated energy transfer to larger aggregates. We show that energy transfer between electronically uncoupled but vibronically resonant donor-acceptor sites does not depend on the intermediate site energy or the number of intermediate sites. The intermediate sites simply mediate electronic coupling such that vibronic coupling along specific promoter modes leads to direct donor-acceptor energy transfer bypassing any intermediate uphill energy transfer steps. We show that interplay between the electronic Hamiltonian and the effective mode transformation partitions the linear vibronic coupling along specific promoter modes to dictate the selectivity of mediated energy transfer, with a vital role of interference between vibronic couplings and multi-particle basis states. Our results suggest a general design principle for enhancing energy transfer through synergistic effects of vibronic resonance and weak mediated electronic coupling, where both effects individually do not promote efficient energy transfer. The effective mode approach proposed here paves a facile route towards four-wavemixing spectroscopy simulations of larger aggregates without severely approximating resonant vibronic coupling. 
\end{abstract}


\maketitle
\section{Introduction}\label{intro}
Impulsively excited vibrational wavepackets often accompany ultrafast electronic energy or charge transfer dynamics as mere spectators and could be understood within the Born-Oppenheimer approximation. However, specific vibrational motions can sometimes strongly couple with electronic motions causing breakdown of the adiabatic framework to drive ultrafast internal conversion between electronic states. Such examples may include initial steps of photosynthesis\cite{JonasARPC2018}, photochemistry of vision\cite{Cerullo2010}, and phase transitions in quantum materials\cite{Cavalleri2016}. Identifying vibrational motions which promote vibronic mixing opens an interesting avenue of driving state selective photochemistry such as inhibiting `promoter' modes to extend excited state lifetime\cite{Paulus2020}, or driving promoter modes to modulating charge transfer in organic crystals\cite{Dawlaty2017,Frontiera2020}, transition metal dichalcogenides\cite{Prezhdo2016} and donor-bridge-acceptor molecules\cite{Weinstein2014}.  

Path integral\cite{MakriARPC2022} and effective modes schemes\cite{Burghardt2005,Burghardt2019} can in principal treat all intramolecular Franck-Condon (FC) vibrational modes of the system on the same footing to provide numerically exact microscopic view of quantum decoherence even for molecular aggregates. While calculations of population transfer rates including the full multidimensional vibrational subspace are now feasible, calculations of spectroscopic signatures of four-wave mixing spectroscopies, which scale as fourth power of the number of basis states may be too expensive still. 

Demonstrations of mode-selective photochemistry suggest that a differentiability between spectroscopically  meaningful spectator and promoter vibrational motions is desirable for guiding synthetic design and optical spectroscopy. Promoter modes correspond to inter- or intramolecular vibrational modes with specific motions, and sometimes definite symmetries, which mix electronic degrees of freedom to dominate the short-time quantum dynamics\cite{Burghardt2005}. Promoter modes are therefore spectroscopically interesting to identify. Effective modes constructed by combining several intramolecular FC vibrations through bilinear couplings\cite{Burghardt2019} may not provide a spectroscopically meaningful distinction between promoter versus spectator vibrational modes.

Several theoretical approaches\cite{Valkunas2014, Schroter2015, Mancal2016} for simulating quantum dynamics and four-wavemixing spectroscopies of molecular aggregates have relied on the above distinction to treat promoter modes explicitly in the system Hamiltonian, while quantum relaxation of singly-excited states due to rest of the bath is treated using quantum Master equations, or symmetric and asymmetric Brownian oscillators when a symmetry based distinction\cite{KitneyHayes2014} between singly-excited electronic states is possible. A numerically exact treatment of the system Hamiltonian of a molecular aggregate with several singly-excited states and vibrational modes of spectroscopic interest requires an exponentially scaling basis set\cite{Tiwari2020}, and becomes computationally impractical for four-wavemixing simulations which further scale to the 4$^{th}$ power of basis set size. This often necessitates a description of vibronic excitons which relies on scaling down vibrational dimensionality of the basis by restricting vibrational excitations on ground electronic states. Such approximations, namely the one-particle approximation\cite{Philpott1969} and numerically similar coherent exciton scattering approximation\cite{Briggs1970}, are very successful in describing linear spectroscopic properties of large molecular aggregates\cite{Briggs2008} and organic thin films\cite{Hestand2018}. However, we have recently shown\cite{Tiwari2020} that specific situations, such as vibronic resonances\cite{Tiwari2013} cause multi-particle basis states to gain substantial oscillator strength. The resulting exciton delocalization, vibrational distortion field and quantum dynamics cannot be described under such approximations. 

The above challenges suggest that theoretical approaches which can identify spectroscopically meaningful promoter vibrational motions, reduce vibrational dimensionality of the multidimensional energy transfer problem, and treat non-adiabatic vibronic coupling exactly can serve as a useful tool to guide molecular design, and spectroscopic experiments and simulations. Early theoretical treatments\cite{Moffit1960,Gouterman1961,Sinanoglu} of energy transfer in a dimer have analyzed the problem in terms of physically motivated tuning and correlation effective modes, akin to longitudinal and totally symmetric deformations of a quantum particle in a 2-dimensional (2D) box potential. For a given FC active intramolecular vibrational mode, $\hat{q}_{A}$ and $\hat{q}_{B}$ on molecules $A$ and $B$ respectively, the dimer energy transfer Hamiltonian is transformed as a sum of separable 1D Hamiltonians, that is, $\hat{H}(\hat{q}_{A}, \hat{q}_{B}) = \hat{H}(\hat{q}_{+}) +\hat{H}(\hat{q}_{-})$. Jonas and co-workers have shown\cite{Tiwari2013, Tiwari2017, Peters2017} that anti-correlated motions along $\hat{q}_-$ tune the singly-excited state energy gaps, and are solely responsible for driving non-adiabatic energy transfer between vibronically resonant states. The $\hat{q}_-$ mode is akin to the tuning vector in conical intersections that defines the direction of Hellmann-Feynman forces\cite{Baer2006} in non-adiabatic transitions. Correlated motions along $\hat{q}_+$ do not tune singly-excited state energy gaps, play no role in vibronic mixing, and can be treated under the adiabatic framework. As far as the role of vibronic coupling in influencing the dynamics and spectroscopic signatures is concerned, the dimensionality of the problem is reduced to $\hat{H}(\hat{q}_{-})$ while still treating vibronic coupling exactly. 

The above analysis of dimer energy transfer in terms of physically intuitive tuning and correlation vibrational modes motivated our earlier work\cite{Patra2021} where an extension of these effective modes to larger molecular aggregates was proposed. Taking the example of a 3-mer, we showed that linear combination of correlation and tuning modes akin to dimer and subsequent Gram-Schmidt orthogonalization yields physically meaningful effective normal modes of the aggregate - a global correlation mode which does not tune any energy gap, a global tuning mode which tunes all nearest-index energy gaps, and a second-nearest index tuning mode. The above transformation preserves the vibrational frequencies of the system and does not yield bilinearly coupled effective modes. Interestingly, expressing the Hamiltonian in terms of the effective modes leads to a sum of 1D Hamiltonians, one along each such effective mode, such that the role of individual effective modes in promoting vibronic mixing can be individually analyzed. The new physical insights gained from this approach suggested an interesting design principle of leveraging vibronic resonances to mediate selective energy transfer to the acceptor in the presence of an intermediate site. 

Here we extend the effective-mode approach and its applications. We elucidate the iterative structure of effective normal modes for successively larger aggregates. Starting from specific 3-mer examples, it is analytically shown that the effective mode transformation partitions the linear vibronic coupling along specific modes with a crucial role played by the electronic Hamiltonian. The design of the electronic Hamiltonian ultimately \textit{selects} the promoter mode by rearranging vibronic couplings to constructively interfere only along specific effective modes. The physical intuition so gained is utilized to generalize idea of selectively mediating energy transfer to $\Lambda$ type systems where multiple intermediate sites and uphill energy transfer steps may be involved. We consider the special case of vibronic resonance between the donor and acceptor excitons, now actively investigated in several photosynthetic proteins\cite{JonasARPC2018}, singlet exciton fission\cite{Rao2016} candidates, and organic polymers\cite{Mohapatra2021}. We show that interference between resonant vibronic couplings along different effective modes is phase-independent and always leads to larger overall couplings. This interference plays a vital role in determining the selectivity of mediated energy transfer between the donor and acceptor by suppressing weaker vibronic couplings with intermediate sites. The intermediate sites simply mediate weak electronic coupling between electronically uncoupled donor-acceptor sites. As long as the donor and acceptor excitons are vibronically resonant, the intermediate uphill energy transfer steps are bypassed to selectively mediate transfer to the acceptor. Our results establish the generality of this design principle for enhancing energy transfer through synergistic effects of vibronic resonance and weak mediated electronic couplings, where both effects by themselves cannot promote efficient energy transfer. Similar mechanisms may be operative in energetically disordered molecular aggregates with a large number of FC active vibrations such as photosynthetic proteins and organic photovoltaic thin films. Our results point to a vital role of multi-particle basis states in describing such mechanisms, and provide a feasible route towards four-wavemixing spectroscopy simulations of larger aggregates without severely approximating resonant vibronic coupling.

The manuscript is organized as follows. Section \ref{theory} formalizes the effective mode approach  highlighting its iterative nature. The section concludes with a physically intuitive picture for the nature of derived effective modes. Section \ref{apps} applies the effective mode formalism to identify promoter modes in general $\Lambda$- or V- type systems along which vibronic coupling strengths are maximized. The section illustrates interesting effects such as role of $\hat{H}_{elec}$ in determining the promoter mode, interference between vibronic couplings along different effective modes, and its role in determining the selectivity of mediate energy transfer. Section \ref{conclusions} presents the conclusions.

\section{Theory} \label{theory}

Our recent work generalized the pairwise tuning and correlation modes of an excitonic dimer to a $N$-mer with total $V$ intramolecular FC active modes on each molecule. The dimer tuning and correlation modes, $\hat{q}_-$ and $\hat{q}_+$ respectively, could be combined to give global tuning and correlation modes, $\hat{Q}^-$ and $\hat{Q}^+$ respectively. A Gram-Schmidt orthogonalization process leads to the residual modes such that the $NV$-dimensional Hamiltonian can be written as a sum of 1D Hamiltonians along the effective modes $\hat{Q}^+$, $\hat{Q}^-$ and $NV-2$ residual modes $\hat{R}$. Below we will sketch the derivation for a $N$-mer, and in the process elucidate the iterative nature of the scheme for successively larger aggregates.

\subsection{Hamiltonian}
Each molecule of the aggregate is assumed to be a two electronic level system with site basis states $\ket{G}$ and $\ket{E}$. The molecules are electronically coupled through Coulomb interactions between their ground to excited state transition dipoles. The resulting electronic basis for the aggregate is constructed from a tensor product of the site basis of each molecule. This results in a set of $N$ singly-excited electronic basis states, where $N=5$ for the case of 5-mer considered here. A singly-excited state $\ket{I}$ denotes the state $\ket{G_{A}G_{B}\dots E_{I}\dots G_{E}}$, where only the $I^{th}$ molecule is electronically excited. The purely electronic part of the singly-excited Hamiltonian of the aggregate is given by $\hat{H}_{elec} = \sum_{I} \big[\epsilon_I \ket{I}\bra{I} + \sum_{L<I} J_{LI}(\ket{L}\bra{I} + \ket{I}\bra{L})\big]$, where the Coulomb coupling matrix element $J_{LI}$ between any two singly-excited electronic states $\ket{L}$ and $\ket{I}$ is a real quantity. $\epsilon_I$ is the electronic site energy for state $\ket{I}$. It is assumed that only the Coulomb integrals contribute to electronic coupling with negligible electron exchange, although the approach described below can be extended to include charge-transfer couplings as well \cite{Patra2021}. Note that no specific spatial arrangement and mutual electronic couplings has been assumed. \\

A common set of $V$ intramolecular vibrational modes are present on the ground and excited electronic state of each molecule, such that $d_{I_j}$ denotes the FC displacement on the singly-excited electronic state of the $I^{th}$ molecule along the $j^{th}$ mode. Note that there is no loss of generality because the set of FC displacements on each molecule can be different. The dimensionless vibrational coordinate for the $j^{th}$ mode on molecule $I$ is denoted by the unit vector operator $\hat{q}_{I_j}$. The corresponding nuclear momentum unit vector operator is denoted by $\hat{p}_{I_j}$. The ground electronic state Hamiltonian of the aggregate is given by $\hat{H}_{G} = \sum_{I}\sum_{j=1}^{V}{\frac{1}{2}\omega_{j}{(\hat{p}_{I_{j}}^2+\hat{q}_{I_{j}}^2)}}$. We can then write the singly-excited electronic state Hamiltonian, $\hat{H}_{N}$ for the case of a $N$-mer, as -- 
\begin{eqnarray} 
{\hat{H}_{N}} = \hat{H}_{elec} + \sum_{I}\big[\hat{H}_G - \sum_{j=1}^{V}\omega_{j}d_{I_j}\hat{q}_j\big]\ket{I}\bra{I}.
\label{eq1}
\end{eqnarray}
The energy is defined in frequency units. Indices $I$ and $L$ run over the molecules $A$ to $E$ of the 5-mer. Linear vibronic coupling in the first term of Eqn.~\ref{eq1} has contributions from FC displacements on the excited state of molecule $I$ along all the intramolecular vibrational coordinates. Note that the vibrational subspace in $\hat{H}_{G}$, and within each electronic subspace in a $N$-mer singly excited Hamiltonian $\hat{H}_N$, has a dimensionality of $N\times V$. Truncating the Hilbert space spanned by the vibrational basis states such that only $n_{vib,g}$ and $n_{vib,e}$ vibrational quanta are allowed on the ground and excited electronic states respectively, the number of basis states in a numerically exact description of energy transfer scale rapidly as $N.(n_{e,vib})^{V}.(n_{g,vib})^{V(N-1)}$. One-particle approximation\cite{Philpott1969,Rashba1965} (1PA), numerically similar to coherent-exciton scattering approximation (CES), where ground electronic state vibrations are restricted to only the lowest vibrational state, can substantially scale down the basis set size. Such approximations or their variants such as 2PA have been successfully used to describe linear absorption and emission lineshapes in vibronic dimers\cite{Briggs1972,Briggs2005}, J- and H- aggregates\cite{Briggs2008} of organic thin films\cite{Hestand2018}, etc. Similar approximations have been employed\cite{Briggs2009,Womick2011, Christensson2012,Thorwart2015,Dean2017,Fleming2020_1} to describe quantum dynamics of vibronic excitons. We have recently shown\cite{Tiwari2020} that basis sets with restricted ground state vibrations may not accurately capture the quantum dynamics arising from vibronic resonances\cite{Tiwari2013}, which are currently an active subject of investigation\cite{Scholes2017} because of the exciting possibility of vibronically enhanced energy and charge delocalization.

\subsection{Effective Normal Modes}\label{normalmodes}
The iterative nature of effective normal modes can be understood through mathematical induction by deriving effective modes for successively larger aggregates starting from a dimer. In our earlier work\cite{Patra2021}, we derived effective modes for a 3-mer. Below we build up from this work to derive a set of effective normal modes for a 5-mer, corresponding to each set of the total V intramolecular vibrational modes per molecule. In doing so, we will highlight the general structure as well as the FC displacements associated with these delocalized normal modes. \\

The singly-excited 5-mer Hamiltonian $\hat{H}_5$ is 5V-dimensional in the starting intramolecular vibrational basis. For any given set $j$ of intramolecular modes with frequency $\omega_j$ on each molecule, we will assume equal FC displacements for simplicity. That is, $d_{I_j} = d_j$ for any molecule $I$. Since only linear transformations are involved in the derivation, such a simplification does not limit the generality of the method. For example, see the derivation of effective normal modes with unequal FC displacements for the case of a 3-mer in Section S1.1. Minor relative differences in vibrational frequencies and FC displacements in disordered aggregates such as a photosynthetic protein will only manifest on longer timescales. \\

Using the definitions in Eqn.~2 and Eqn.~3 of ref.\cite{Patra2021} for the case of a $N$-mer, the $j^{th}$ set of global tuning and the correlation vectors, $\textbf{G}_j$ and $\textbf{C}_j$ respectively, are linear combinations of all the pairwise tuning ($\textbf{g}^{p_n}$) and correlation vectors ($\textbf{c}^{p}$), and written compactly as --
\begin{eqnarray}
\textbf{G}_j.\mathbf{\hat{q}} &=& \textbf{m}_j^{p_n}.(\hat{\textbf{g}}_j^{p_{n}}.\hat{\textbf{q}})\\ \nonumber
\textbf{C}_j.\mathbf{\hat{q}} &=& \textbf{n}_j^{p}.(\hat{\textbf{c}}_j^{p}.\hat{\textbf{q}})
\label{eq2}     
\end{eqnarray}
Here index $p_n$ denotes the $N-1$ nearest-neighbor combinations, and index $p$ denotes all the possible $N(N-1)/2$ pairs of sites. Note that the usage of term nearest-neighbor should be interpreted in terms of nearest indices and not in terms of spatial proximity. For the case of global tuning vector for a 5-mer, $\textbf{m}_j^{p_n} = (m_j^{AB}, m_j^{BC},m_j^{CD},m_j^{DE})$. For the case of global correlation vector, $\textbf{n}_j^p = (n_j^{AB}, n_j^{AC},n_j^{AD}\dots n_j^{DE})$. Both denote a vector of unknown coefficients that dictates the weightage of pairwise tuning (correlation) modes towards the global tuning (correlation) mode. Eqn.~2 represents a dot product of this `weighting' vector with another vector formed by the contributions of pairwise tuning(correlation) motions towards adjusting the singly-excited state energies of the two sites that form the pair, through linear vibronic coupling. Patra et al.\cite{Patra2021} show that the unknown coefficients can be obtained by imposing intuitive geometric constraints. For example, the global tuning vector $\textbf{G}_j$ is orthogonal to all second-nearest neighbor energy gap tuning vectors. Similarly, the global correlation vector $\textbf{C}_j$ is orthogonal to all the pairwise tuning vectors in the system, as well as the global tuning vector $\textbf{G}_j$. Imposing such constraints to determine the unknown coefficients, and normalizing the resulting vibrational modes yields global tuning and correlation modes for the 5-mer, $\hat{Q}_j^+$ and $\hat{Q}_j^-$ respectively.\\

\begin{figure*}[h!]
	\centering
	\includegraphics[width=3 in]{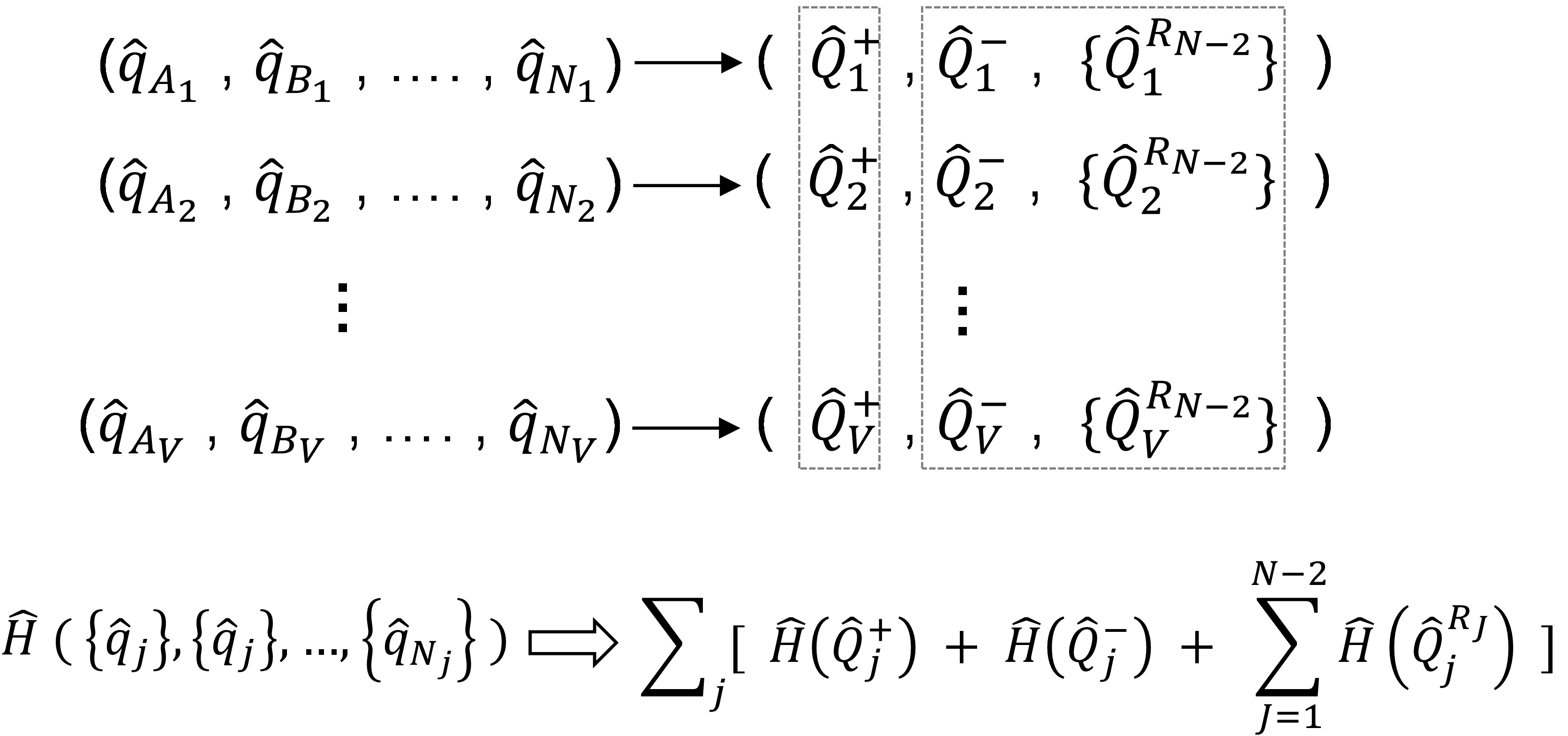}
	\caption{\footnotesize Effective Mode Transformation. Effective normal modes for an aggregate of $N$ molecules each having $V$ intramolecular FC modes. Each set of total $V$ set of modes corresponds to a particular set of $N$ effective modes. The effective modes $\hat{Q}_j^+$ do not tune any energy gaps and are mere spectators in vibronically enhanced energy transfer. Other effective modes can either tune energy gaps, nearest neighbor energy gaps in case of $\hat{Q}^-$ and non-nearest neighbor in case of $N-2$ residual modes $\hat{Q}^{R}$, to affect short-term non-adiabatic dynamics depending on how strongly they couple different electronic degrees of freedom. Using the effective mode transformation, the $NV$ dimensional energy transfer Hamiltonian can be written as a sum of $N \times V$ 1D Hamiltonians, one along each effective mode.  
	}
	\label{fig:fig1}
\end{figure*}

As explained by Patra et al.\cite{Patra2021}, the remaining $N-2$ residual modes are determined by imposing the constraint in each of the $N$ electronic domains that the total linear vibronic coupling arising from the $j^{th}$ set of intramolecular vibrational modes remains the same in the delocalized vibrational basis. Note that this results in an overdetermined system of $N-2$ unknowns with $N$ constraints, yielding \textit{linearly dependent} residual modes. This leads to a certain flexibility in designing the linearly independent residual modes using Gram-Schmidt orthogonalization (Section \ref{iterative}). After orthonormalization, the resulting set of effective normal modes for the 5-mer can be expressed as a linear transformation ${\textbf{Q}}_j = U_{5 \times 5}^{-1}{\textbf{q}}_j$, where $\textbf{Q}_j = (\hat{Q}_j^+, \hat{Q}_j^-, \hat{Q}_j^{ACE}, \hat{Q}_j^{AE}, \hat{Q}_j^{BD})$ is a column vector of the $j^{th}$ set of delocalized effective modes for the 5-mer. The superscripts on the three residual modes indicate the diabatic electronic sites on which the residual effective mode has FC displacements. The physical meaning and choice of these residual modes will be discussed in Section \ref{iterative}. $\textbf{q}_j = (\hat{q}_{A_j}, \hat{q}_{B_j}, \hat{q}_{C_j}, \hat{q}_{D_j}, \hat{q}_{E_j})$, is a column vector of the $j^{th}$ set of intramolecular vibrational modes. The orthogonal transformation matrix $U_{5 \times 5}$, shown in Eqn.~S12, transforms the $j^{th}$ set of intramolecular vibrational modes into delocalized effective normal modes of the aggregate. A similar transformation $U_{3 \times 3}$ is shown in Section S1.1. Two crucial points that highlight the generality of the approach should be noted here -- 1. No $j$ subscript on the transformation $U$ is intentional because the same orthogonal transformation is valid for any set $j$ in the total $V$ sets of intramolecular modes. This is highlighted schematically in Figure \ref{fig:fig1} and discussed in Section \ref{iterative}. 2. The coefficients of orthogonal transformation $U$ also determine the set of FC displacements along the corresponding $j^{th}$ set of delocalized effective modes in $\textbf{Q}_j$. The set FC displacements for the 3- and 5-mer are tabulated in Tables S1 and S2 respectively. The derivation of 5-mer effective modes is detailed in Section S1.2.

Overall, the orthogonal transformation $U_{N \times N}$ between intramolecular and delocalized effective modes allows one to express the linear vibronic coupling terms of the aggregate Hamiltonian $\hat{H}_N$ in Eqn. \ref{eq1} in terms of the delocalized effective normal modes. As explained schematically in Figure \ref{fig:fig1}, this amounts to writing the $NV$-dimensional Hamiltonian in Eqn.\ref{eq1} as a summation of $N \times V$ 1D Hamiltonians, one along each effective mode. The 1D Hamiltonians for the 5-mer are shown in Eqns.~S13. From this simplified structure, the set of global correlation modes $\{\hat{Q}_j^+\}_{j=1}^V$, akin to totally symmetric deformation of a $N$-dimensional box, do not tune any singly-excited state energy gaps in the aggregate, and play no role in mixing the electronic domains in $\hat{H}_{elec}$. As shown in our previous work\cite{Patra2021} for the case of a 3-mer, in order to ascertain the role of remaining $N(V-1)$ modes in promoting vibronically enhanced energy transfer, significantly faster calculations along the reduced 1D Hamiltonians can be carried out, such that effective modes which promote vibronic mixing versus those which are mere spectators could be differentiated. While extensive effective-mode schemes\cite{Burghardt2005} which can treat all the intramolecular vibrational modes on the same footing have been demonstrated for excitonically coupled aggregates\cite{Burghardt2019}, the unique aspects of the current approach, although may be only limited to a few spectroscopically observed vibrations of interest, is that the motions along these effective vibrational modes are physically interpretable, and their frequencies directly relate to spectroscopic observables. The effective mode structure proposed here is not bilinearly coupled (orthogonal) and preserves the vibrational frequencies of the system. Thus a reduction in the dimensionality of the vibrational sub-space is made possible by identifying promoter versus spectator vibrational motions. We have demonstrated this aspect in our original communication\cite{Patra2021}, and will further extend that in the current manuscript. Further, as opposed to approximations\cite{Tiwari2020} relying on multi-particle basis set truncation, the non-adiabatic vibronic mixing and the resulting enhancement of energy or charge transfer is treated numerically exactly in the effective mode transformation.  

\subsection{Iterative Nature of Physically Meaningful Effective Modes}\label{iterative}

Because the design of the effective normal modes preserves the vibrational frequencies of the system with no bilinear couplings between modes, an iterative structure for the effective modes of successively larger aggregates becomes possible. As mentioned above, the $N$ electronic domains for $N-2$ (to be determined) residual modes lead to a certain flexibility in constructing the residual effective modes through Gram-Schmidt orthogonalization. We start with the case of a 3-mer, where the global correlation mode $\hat{Q}_{j}^+$ does not tune any singly-excited energy gap, and the global tuning mode $\hat{Q}_{j}^-$ tunes all the nearest-neighbor energy gaps. The only choice of residual mode turns out to be the mode $\hat{Q}_j^{AC}$ which tunes the remaining $A-C$ energy gap, or the second nearest-neighbor energy gap in the system. The superscript indicates the electronic domains with non-zero FC displacements along the effective mode. 

\begin{figure*}[h!]
	\centering
	\includegraphics[width=4 in]{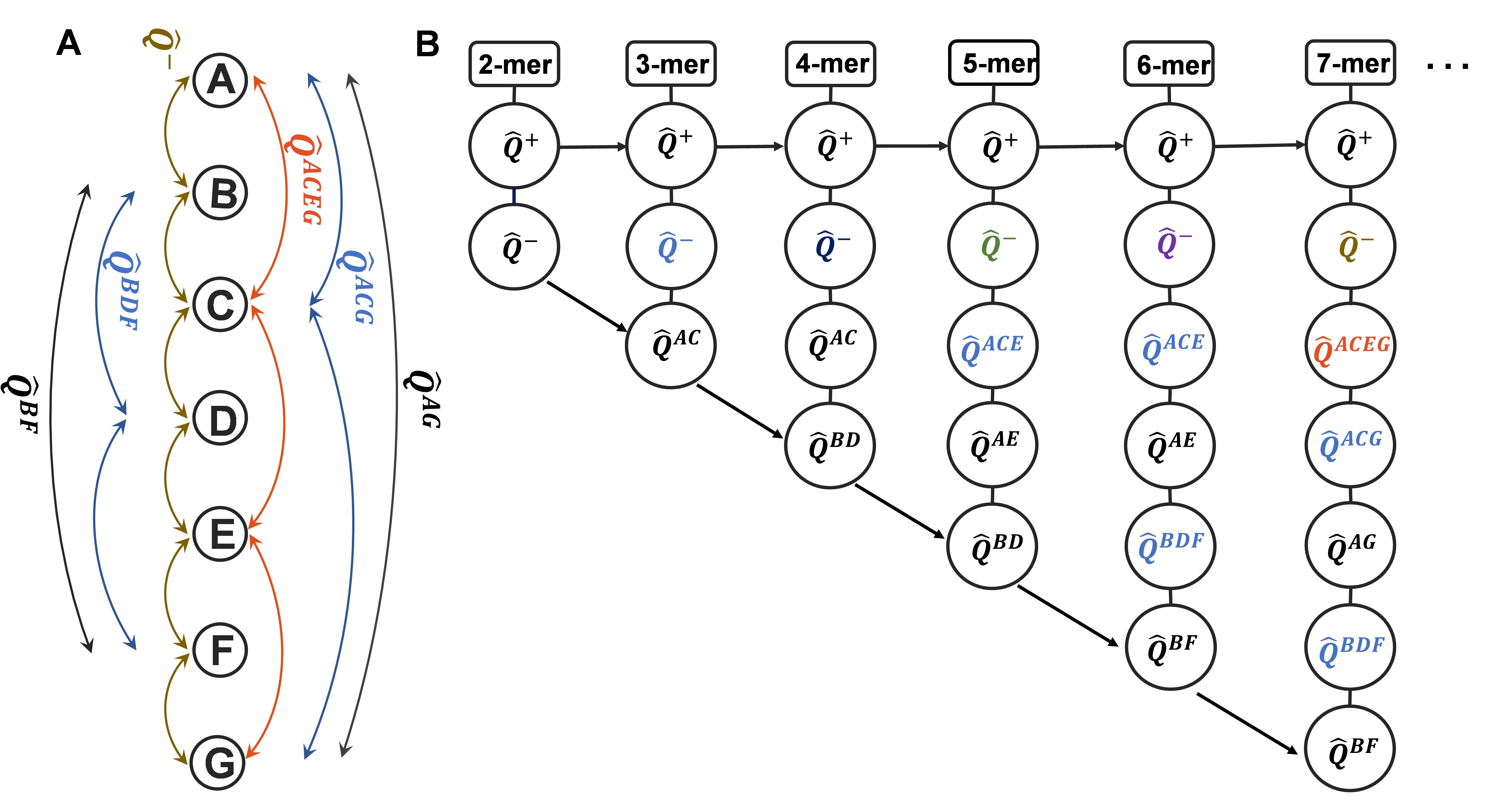}
	\caption{\footnotesize  (a) Schematic depiction of global tuning and residual effective modes for the case of a $N$-mer. The superscript on the residual modes indicate the electronic domains with non-zero FC displacements. For any given effective mode, each curved line connecting two electronic domains corresponds to the energy gap tuning vector between the two domains, such that pairwise combinations of all such vectors gives rise to the overall effective mode. This set of effective normal modes correspond to each of the total $V$ sets of intramolecular vibrations. The subscript $j$ to indicate the $j^{th}$ set is therefore suppressed here for clarity. (b) Iterative structure of the effective normal modes for successively larger aggregates. The form of residual modes is based on the schematic shown in panel a. The global correlation mode, $\hat{Q}^+$ has a FC displacement of $\frac{1}{\sqrt{N}}$ in each of the $N$ electronic domains. The global tuning mode $\hat{Q}^-$ and the residual modes are color coded to map effective modes which have identical sets of FC displacements. The electronic domains with non-zero FC displacements along residual modes are shown in superscripts. The map indicates that several effective normal modes and FC displacements for a general $N$-mer are equivalent to effective modes for smaller aggregates. For example, for a 7-mer, the only unknowns are the new colors, that is, the FC displacements along $\hat{Q}^-$, and along the new residual mode $\hat{Q}^{ACEG}$.      
	}
	\label{fig:fig2}
\end{figure*}

A general schematic for choosing the residual modes for a $N$-mer is shown in Figure \ref{fig:fig2}A. Applying this schematic to a 5-mer, the residual effective modes become $\hat{Q}_j^{ACE}, \hat{Q}_j^{AE}, \hat{Q}_j^{BD}$. The choice of residual effective modes fixes the electronic domains in which the residual mode has zero and non-zero FC displacements.   Other equivalent choices of residual modes are also possible and simply lead to residual modes with the same set of FC displacements but on different electronic domains. This is shown in Section S1.3. The knowledge of effective modes and corresponding FC displacements for smaller aggregates, such as a 3-mer, implies that the unknown FC displacements along the residual effective modes for a larger aggregate, such as a 5-mer, are already known. This is depicted in Figure \ref{fig:fig2}B in the form of color-coding, where modes of similar colors have the same set of FC displacements with the electronic domains denoted by the superscripts. The FC displacement along the global correlation mode $\hat{Q}_j^+$ is also known -- $d_j/\sqrt{N}$ in each of the $N$ electronic domains. An important point to also recall here is that for a given $N$-mer, the transformation $U_{N\times N}$ holds for any set $j$ of the total $V$ sets of intramolecular FC modes. \textit{Thus for any larger aggregate the only unknowns are the FC displacements along the global tuning mode $\hat{Q}_j^-$ and along the new residual modes, such as $\hat{Q}_j^{ACEG}$ for the case of a 7-mer.}   

\begin{figure*}[h!]
	\centering
	\includegraphics[width=3 in]{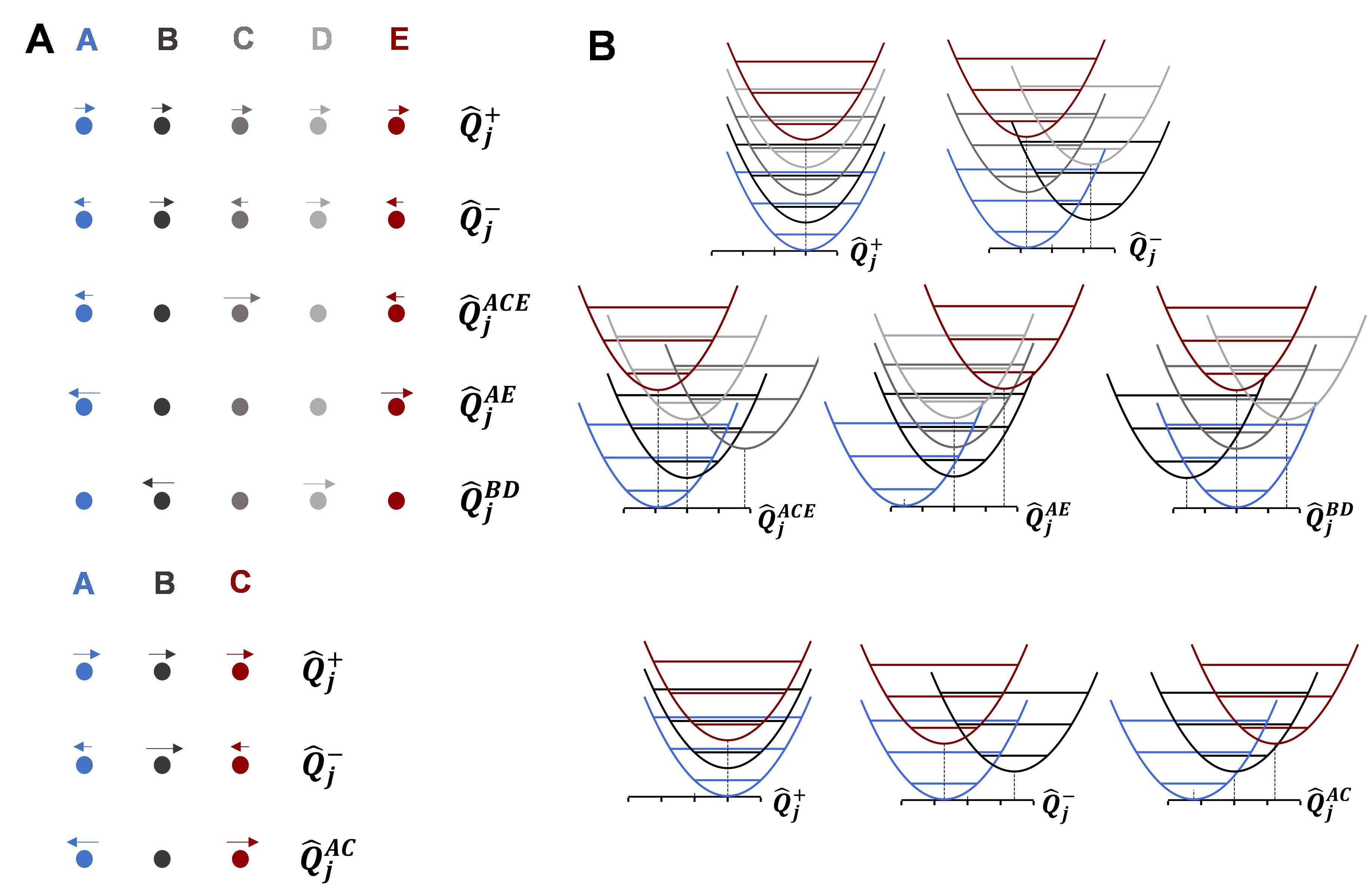}
	\caption{\footnotesize (\textbf{A}) FC displacement patterns on singly-excited diabatic electronic sites along the effective normal modes. The sites are color coded to map to the diabatic potentials in panel B. The top and bottom panels correspond to 5- and 3-mer respectively. The direction and size of the arrow on each site maps to the sign and magnitude of FC displacement on the diabatic electronic site, respectively. The FC displacements are shown in Table S3. (\textbf{B}) 1D slices of singly-excited diabatic electronic potentials for a general electronic Hamiltonian $\hat{H}_{elec}$ plotted along the effective normal modes for a 5-mer (top) and a 3-mer (bottom). The classical turning points 0, $\pm$1, $\pm$2 are marked on the horizontal axis for reference, and the respective FC displacements are shown on the same scale. Dashed vertical lines serve as a guide to connect the minima of each curve to the horizontal nuclear coordinate, and highlight the FC displacements along each effective mode. For ease of visualization, the intramolecular FC displacement on each site is chosen to be $d = \sqrt{N}$ for a $N$-mer. For weakly coupled FC vibrations considered here, the effective FC displacements in the diabatic site basis will be a small fraction of the classical turning points of the zero-point level.}
	\label{fig:fig3}
\end{figure*}

Table S3 lists the FC displacements in the diabatic site basis along all the delocalized effective modes shown in Fig.~\ref{fig:fig2}B.  The derivation of FC displacements in Table S3 follows the above formalism and discussed in more detail in Section S1. To emphasize the iterative nature of effective modes, Figure \ref{fig:fig3} illustrates the pattern of FC displacements and the diabatic site potentials along the $j^{th}$ set of effective modes for a 3-mer and 5-mer with a general $\hat{H}_{elec}$. Panel A plots the pattern of FC displacements along each effective mode in the diabatic site basis. The direction and size of arrow on top of each site map to the signs and magnitudes of FC displacements on the respective sites. Panel B plots the diabatic site potentials with the FC displacements along the effective modes shown on scale with the classical turning points. All diabatic sites have equal FC displacements along $\hat{Q}^+$ modes, such that motions along $\hat{Q}^+$ cause no change in relative energy gaps. In contrast, motions along $\hat{Q}^-$ tune relative energy gaps between nearest neighbor excitons, but do not tune the second-nearest neighbor energy gaps. In the same fashion, motions along effective modes such as $\hat{Q}^{ACE}$ do not tune the relative $B-E$ energy gap. The pattern of FC displacements, including the sign and magnitude, for the 5-mer effective modes $\hat{Q}^{AC}$ and $\hat{Q}^{BD}$ are identical to those along the 3-mer effective modes $\hat{Q}^{AC}$. Similarly the 5-mer effective mode $\hat{Q}^{ACE}$ has FC displacements identical to the 3-mer global tuning mode $\hat{Q}^{-}$.  \\

From Figs.~\ref{fig:fig2}-\ref{fig:fig3} and Table S3, it is seen that the transformation to effective normal modes essentially partitions the linear vibronic couplings arising from intramolecular FC displacements unequally along the delocalized effective modes. The effective modes that are delocalized over more number of sites contribute to smaller vibronic couplings. For example, compare FC displacements  for a 2-mer versus a 7-mer, corresponding to the global tuning mode $\hat{Q}^-$ in Table S3.  It may be expected that more delocalized effective modes influence the vibronic dynamics on longer timescales on account of reduced vibronic coupling. However, Section \ref{promoter} and Section \ref{select} illustrate the crucial role of the purely electronic Hamiltonian $\hat{H}_{elec}$ in ultimately determining the dominant promoter mode by rotating\cite{Peters2017,Patra2021} the linear vibronic couplings to strongly couple only specific electronic domains. Section \ref{interfere} discusses that constructive or destructive interference between vibronic coupling matrix elements\cite{Patra2021, Tiwari2018, Tiwari2020} arising from different electronic sites, effective modes or vibrational frequencies are secondary effects that also dictate the total vibronic mixing along the promoter effective modes.  

\section{Applications} \label{apps}
In the following sections, the above formalism will be applied to generalize the concept of mediated energy transfer proposed\cite{Patra2021} in our previous work. Using examples of $\Lambda$- or V-type aggregates, we will apply the effective mode approach to identify the dominant promoter modes which maximize vibronic mixing between excitons. The knowledge gained will be used to illustrate interesting effects such as the role of $\hat{H}_{elec}$ in determining the promoter mode, interference between vibronic couplings in determining the selectivity of mediated energy transfer, and the synergy between weak mediated electronic couplings and vibronic resonances in affecting direct vibronically enhanced energy transfer skipping intermediate uphill energy transfer steps.

\subsection{Promoter Modes in a 3-mer}\label{promoter}
We will use the formalism described in Section \ref{theory} to identify the promoter modes in a 3-mer system along which vibronic mixing between excitons is maximized. As we will see, this understanding is necessary in order to generalize the idea of mediated energy transfer to larger aggregates, and is made possible through the effective-mode approach presented here.
Section S1 derives the $U_{3 \times 3}$ transformation for a 3-mer with unequal intramolecular FC displacements $d_{A}$, $d_{B}$, $d_{C}$ along any given set of intramolecular vibrational modes on sites $A$, $B$ and $C$, respectively. The transformation $U_{3 \times 3}$ connects the set of intramolecular vibrational modes to the corresponding set of effective normal modes $\hat{Q}^+$, $\hat{Q}^-$ and $\hat{Q}^{AC}$ as --
\begin{equation}
\left(%
\begin{array}{c}
\hat{q}_A \\
\hat{q}_B \\
\hat{q}_C \\
\end{array}%
\right)=\left(%
\begin{array}{ccc}
\frac{1}{D_3 d_A} & -\frac{1}{D_3 D_{AC} d_A d_B} & -\frac{d_A}{\sqrt{d_A^2 + d_C^2}} \\
\frac{1}{D_3 d_B} & \frac{D_{AC}}{D_3} & 0 \\
\frac{1}{D_3 d_C} & -\frac{1}{D_3 D_{AC} d_C d_B} & +\frac{d_C}{\sqrt{d_A^2 + d_C^2}}  \\
\end{array}%
\right)\left(%
\begin{array}{c}
\hat{Q}^{+} \\
\hat{Q}^{-} \\
\hat{Q}^{AC} \\
\end{array}%
\right)\label{eq3}
\end{equation}

Here $D_3 = \sqrt{\frac{1}{d_A^2} + \frac{1}{d_B^2} + \frac{1}{d_C^2}}$ and $D_{AC} = \sqrt{\frac{1}{d_A^2} + \frac{1}{d_C^2}}$. Note that the subscript corresponding to $j^{th}$ set of FC vibrational modes has been dropped in Eqn.~\ref{eq3} for brevity. As shown in Section S1.1, the FC displacements along the effective modes on each electronic domain can be derived from $U_{3 \times 3}$ in a straightforward manner. Table S1 lists these FC displacements. Expressing the vibrational part of the Hamiltonian in Eqn.~\ref{eq1} in terms of the effective modes, the vibrational Hamiltonian is split into three 1D Hamiltonians corresponding to each set $j$ of intramolecular vibrational modes --
\begin{eqnarray}
\hat{H}_3 &=& \hat{H}_{elec} + \sum_{j}\hat{H}_3(\hat{p}_{A_j},\hat{q}_{A_j},\hat{p}_{B_j},\hat{q}_{B_j},\hat{p}_{C_j},\hat{q}_{C_j}) \nonumber \\
&=& \hat{H}_{elec} + \sum_{j}\hat{H}(\hat{P}_{j}^{+},\hat{Q}_{j}^{+}) + \hat{H}(\hat{P}_{j}^{-},\hat{Q}_{j}^{-}) + \hat{H}(\hat{P}_{j}^{AC},\hat{Q}_{j}^{AC}).
\label{eq4}
\end{eqnarray}
For any given set of modes, the linear vibronic coupling part of the 3-mer Hamiltonian, $\hat{H}_3^{LV}$, can then be expressed as a sum of vibronic coupling Hamiltonians along individual 3-mer effective modes --
\begin{eqnarray}
	\hat{{H}}_3^{LV} &=&\left(%
	\scalemath{0.8}{
		\begin{array}{ccc}
			-{\omega\hat{Q}^+}{d^+} & 0 & 0 \\
			0 & -{\omega\hat{Q}^+}{d^+} & 0 \\
			0 & 0 & -{\omega\hat{Q}^+}{d^+} \\
		\end{array}%
	}
	\right)\nonumber\\&&+
	\left(%
	\scalemath{0.8}{
		\begin{array}{ccc}
			{\omega\hat{Q}^-}{d_{A}^-} & 0 & 0 \\
			0 & -{\omega\hat{Q}^-}{d_{B}^-} & 0 \\
			0 & 0 & {\omega\hat{Q}^-}{d_{C}^-} \\
		\end{array}%
	}
	\right)+\left(%
	\scalemath{0.8}{
		\begin{array}{ccc}
			{\omega\hat{Q}^{AC}}{d^{AC}_A} & 0 & 0 \\
			0 & 0 & 0 \\
			0 & 0 & -\omega \hat{Q}^{AC}{d^{AC}_{C}} \\
		\end{array}
	}
	\right)
	\label{eq5}
\end{eqnarray}
$d_A^{AC}$ and $d_C^{AC}$ are FC displacements along the $\hat{Q}^{AC}$ effective mode. These are defined as $d_A^{AC} = \frac{d_A^2}{\sqrt{d_A^2 + d_C^2}}$, and $d_C^{AC} = \frac{d_C^2}{\sqrt{d_A^2 + d_C^2}}$. As seen from Eqn.~\ref{eq5}, only $\hat{Q}^-$ and $\hat{Q}^{AC}$ dependent Hamiltonians tune the diabatic site energy gaps and responsible for mixing electronic domains in $\hat{H}_{elec}$. The extent to which $\hat{Q}^-$ and $\hat{Q}^{AC}$ promote vibronic mixing depends crucially on the structure of $\hat{H}_{elec}$. This point is illustrated below.

\begin{figure*}[h!]
	\centering
	\includegraphics[width=3.25 in]{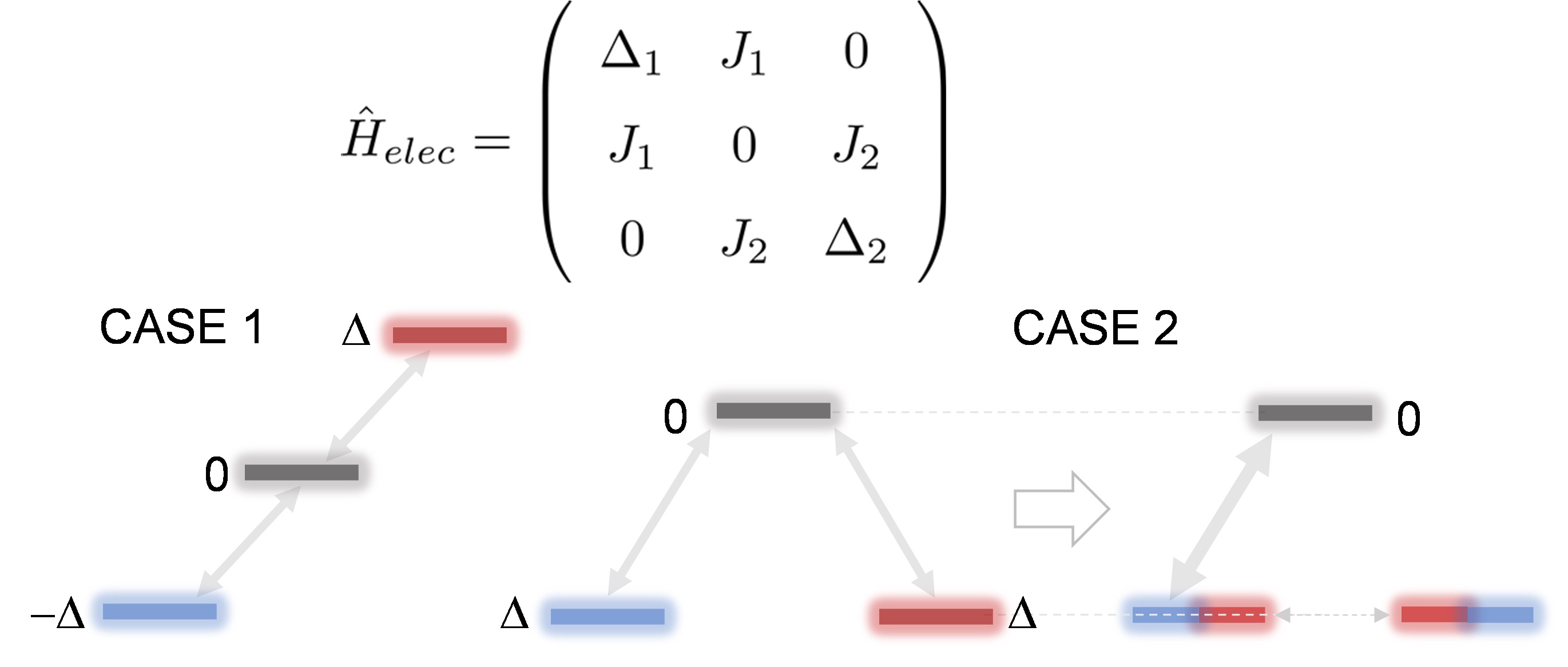}
	\caption{\footnotesize A general 3-mer singly-excited Hamiltonian in the electronic site basis. The electronic domains are arranged as $A-B-C$, where $A$ and $C$ are donor and acceptor electronic states respectively. The electronic energy of the intermediate site is offsetted to be the zero of energy. Specific choices of $\Delta_{1,2}$ leads to `downhill' (Case 1) versus a V- or $\Lambda$-type 3-mer system (Case 2). For \textit{symmetric} V- and $\Lambda$-system site energies $\Delta_1 = \Delta_2 = \Delta$. Eqn.~S15 shows that an electronic transformation modifies the picture such that only one of the resulting $\pm$ linear combination states couples strongly to the intermediate state. The stronger coupling is denoted by bold arrows. The $\pm$ linear combinations of the donor-acceptor sites are denoted as mixed colors.  In case of a more general asymmetric V- or $\Lambda$- system, the $\pm$ linear combinations are still weakly coupled. This weak coupling, zero for the symmetric case, is denoted as thin arrow between the $\pm$ states. The energy levels resulting after this intermediate transformation are shown for the case of equal electronic couplings, that is, $J_1=J_2=J$, although Eqn.~S15 describes this more generally.}
	\label{fig:fig4}
\end{figure*}

The singly-excited 3-mer electronic Hamiltonian is shown in Figure \ref{fig:fig4}, and setup such that the donor and acceptor sites are not directly coupled. For both the cases, the (gray) intermediate site mediates the electronic coupling between the donor and acceptor. The choice of $\Delta_{1,2}$ and electronic couplings $J_{1,2}$ is such that the mediated donor-acceptor electronic coupling is very weak with minimal direct population transfer between the sites. This Hamiltonian can be extended to two general cases depending on site energies $\Delta_1$ and $\Delta_2$. For case 1 both energy transfer steps are downhill, whereas case 2 is a general V-type or a $\Lambda$-type system with a combination of uphill and downhill energy transfer steps for donor to acceptor energy transfer. \\

In general each of the 3-mer molecules contain several FC-active vibrational modes. However, ultrafast experiments on a variety of systems suggest\cite{Dawlaty2017,Frontiera2020,Cavalleri2016,Mathies2005,Paulus2020,Scholes2017} a role for specific vibrational modes in driving ultrafast internal conversion between excited electronic states. Assuming that a particular vibrational mode of spectroscopic interest is observed in a 3-mer system, we want to identify the effective modes which maximally promote vibronic mixing between the weakly coupled donor-acceptor sites. Although the 3-mer toy model used here is rather simple, its merit lies in the ability to \textit{analytically} understand the interplay of electronic Hamiltonian and linear vibronic coupling along the effective modes in eventually determining which vibrational motions act as promoter modes versus motions which do not mix electronic degrees of freedom. The effective-mode approach presented here makes this possible while also treating vibrational-electronic mixing numerically exactly. As we will demonstrate, the physical intuition gained from the analysis of these simple systems is quite useful when thinking about mediated energy transfer in larger aggregates.

In order to understand the interplay of promoter and spectator modes for the `downhill' case 1, we will start with the specific choice of electronic site energies $-\Delta_1 = \Delta_2 = \Delta$ and couplings $J_1=J_2=J$ such that the resulting Hamiltonian is analytically diagonalizable. The zero of energy is chosen to be on the intermediate site. The diagonalizing transformation is given by Eqn.~S16. Section S2.1 applies this transformation on the linear vibronic coupling part of the Hamiltonian. Since $\hat{H}_3^{LV}$ is separable along 1D Hamiltonians, the contributions of $\hat{Q}^-$ and $\hat{Q}^{AC}$ towards vibronic mixing can be individually analyzed. From Eqns. S18-S19, it is seen that the vibronic coupling matrix element along $\hat{Q}^-$ which mixes the donor-acceptor electronic domains is given by $\omega \hat{Q}^-\frac{1}{2}\big[\frac{d_A^2d_C^2}{d_A^2+d_C^2} + d_B^2\big]^{1/2}\sin[2](2\theta)$.  $\theta$ determines the electronic mixing among the 3-mer sites and given by Eqn.~S17. This form of the vibronic coupling implies \textit{direct} vibronic mixing between donor-acceptor sites is promoted by $\hat{Q}^-$ even if there was no direct electronic coupling between the sites. This is so because the effective mode transformation and $\hat{H}_{elec}$ rearrange the intramolecular FC displacements interfere constructively along the global tuning mode $\hat{Q}^-$. In contrast, as shown in Eqn.~S21, along $\hat{Q}^{AC}$, the corresponding vibronic coupling matrix element is $\omega \hat{Q}^{AC}\frac{(d_A^{2} - d_C^{2})}{\sqrt{d_A^{2} + d_C^{2}}}\frac{\sin[2](2\theta)}{4}$. In this case, the vibronic coupling is diminished because it depends only on the \textit{difference} of intramolecular FC displacements, which becomes negligible for case of identical molecules in an aggregate. Thus, for 'downhill' 3-mer (case 1), motions along the global tuning mode $\hat{Q}^{-}$ maximally promote vibronic mixing, while vibrational motions along $\hat{Q}^{AC}$ are spectators in the process. This point was also shown in our earlier work\cite{Patra2021}.

For case 2, we will start with a `symmetric' $\Lambda$-system with electronic site energies $\Delta_1 = \Delta_2 = -\Delta$ and couplings $J_1=J_2=J$. The intermediate site energy is again chosen to be the zero of energy. The choice of V- or $\Lambda$-type system depends on whether the intermediate site is above or below the donor-acceptor electronic energies. Section S2.2 shows that a symmetric $\Lambda$ system can be analytically diagonalized. Upon applying this transformation on the $\hat{Q}^-$ dependent linear vibronic coupling Hamiltonian, the vibronic coupling matrix element between the donor-acceptor electronic domains becomes zero because of \textit{exact} cancellations between intramolecular FC displacements along $\hat{Q}^-$ in the donor-acceptor domains. This is shown in Eqn.~S27. In contrast, as seen in Eqn.~S28, the corresponding vibronic coupling matrix element along the $\hat{Q}^{AC}$ mode depends on the \textit{sum} of intramolecular FC displacements as $-\omega \hat{Q}^{AC}\sqrt{(d_A^{2} + d_C^{2})}\frac{\cos(\theta)}{2}$. \textit{Thus the roles of promoter versus spectator modes are reversed between `downhill' versus symmetric V or $\Lambda$ 3-mer.}

The preceding analysis of the switching roles of `promoter' and `spectator' modes hints at a tempting possibility of engineering electronic Hamiltonians to select which vibrational motions can drive vibronic mixing. It should be emphasized here that the above physical insights are made possible only upon the effective mode transformation from intramolecular to effective normal modes which partitions the linear vibronic coupling from intramolecular vibrations along various effective modes. The electronic Hamiltonian further rearranges these vibronic couplings to interfere constructively only along specific effective modes. As will be illustrated in Section \ref{met}, the resulting vibronic couplings can be leveraged to mediated selective energy only between electronic domains where exciton energy gaps are resonant with vibrational frequencies.

For the more general V- or $\Lambda$-type case, site energies $\Delta_{1,2}$ and couplings $J_{1,2}$ are not equal, and the resulting electronic Hamiltonian is not analytically diagonalizable in a straightforward manner. However an intermediate transformation in Eqn.~S15 shows that the general system can be transformed to a new basis of positive and negative linear combinations of donor-acceptor electronic sites weighted by the mixing angle $\eta = \arctan(J_2/J_1)$. For the general V- or $\Lambda$- 3-mer, we will consider the case where the donor site is strongly coupled to the intermediate site, that is, $J_2 >> J_1$. This implies that the donor-acceptor mixing resulting in $\pm$ linear combination is such that $+$ combination has dominantly acceptor site character and vice versa. The resulting donor-acceptor energy transfer mediated purely by electronic coupling is weak and results in negligible donor to acceptor energy transfer.. The alternative choice of acceptor strongly coupled to the intermediate site is a straightforward extension of the analysis presented here. In the rotated electronic basis, the $\pm$ combinations are now mutually coupled with strength $\frac{\Delta_2-\Delta_1}{2}\sin(2\eta)$, and the -ve combination is no longer coupled to the intermediate site. The corresponding electronic energy gap is $(\Delta_2-\Delta_1)\cos(2\eta)$. The +ve combination is now coupled to the intermediate site with \textit{increased} coupling strength $J_1\cos(\eta) + J_2\sin(\eta)$. Note that this increase is independent of the relative sign of electronic couplings. Figure \ref{fig:fig4} shows the effect of this intermediate transformation for the special case of $J_2 = J_1 = J$.   

\subsection{Mediated Energy Transfer via Vibronic Resonance}\label{met}
In the general $\Lambda$ 3-mer described above, direct donor-acceptor electronic mixing is weak. Therefore any enhancement of electronic energy transfer will require mixing between electronic and vibrational degrees of freedom to either overcome or \textit{bypass} any intermediate uphill steps. The following discussion considers a special case of vibronic mixing which exploits a resonance\cite{Tiwari2013} between excitonic energy gap and a quanta of vibration on the acceptor exciton to enhance energy transfer. This vibronically enhanced transfer is enabled by strong mixing of the donor exciton with the resonantly `selected' vibration on the acceptor.

We consider an excitonically coupled 3-mer  with a specific vibrational mode which participates in vibronic resonance. Even for one specific FC vibrational mode per molecule, the dimensionality scales as $3 \times n_{vib}^{3}$.  However, the above understanding of symmetric `downhill' and $\Lambda$ cases suggests a dominant role for only the $\hat{Q}^{AC}$ effective mode reducing it to a 1D problem to a good approximation.

\begin{figure*}[h!]
	\centering
	\includegraphics[width=3.25 in]{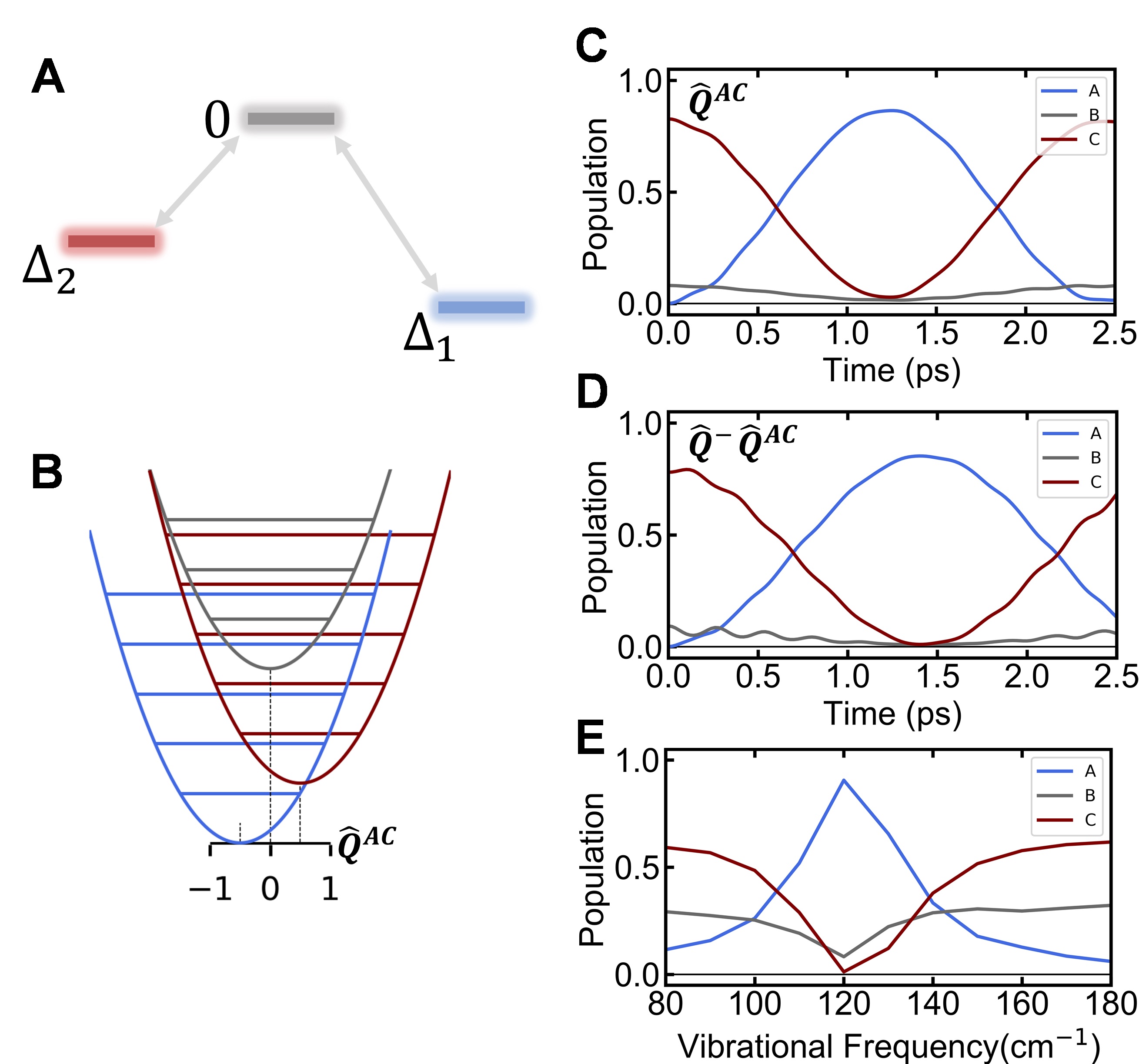}
	\caption{\footnotesize (A) The general $\Lambda$- Hamiltonian. The donor and acceptor site energies are $\Delta_1 = -430$  cm$^{-1}$ and $\Delta_2 = -280 $ cm$^{-1}$, respectively. The electronic couplings are $J_1 =  $40 cm$^{-1}$ and $J_2 = $100 cm$^{-1}$. The intermediate site energy is offsetted to be the zero of energy. (B) Corresponding electronic potential energy curves in the diabatic site basis along the dominant effective mode $\hat{Q}^{AC}$. The diabatic potentials are coupled through the electronic Hamiltonian. The donor, acceptor and intermediate excitons are dominantly the same character as electronic sites. The FC displacements are listed in Table S1. The intramolecular FC displacements are such that $d_A = d_C = \sqrt{0.5}$ and the FC displacement on the intermediate site $d_B = 0$. The FC displacements are shown on scale with the $\pm$1 classical turning points. The vibrational frequency $\omega = 125$ cm$^{-1}$ is chosen to be resonant with the excitonic energy gap between the donor and acceptor excitons. (B) Population dynamics of the $\Lambda$-Hamiltonian after exciting a superposition of vibronically mixed donor and acceptor states upto the 1$^{st}$ resonant manifold. This is a 1D calculation along the dominant promoter mode $\hat{Q}^{AC}$. (D) Population dynamics along both the effective modes $\hat{Q}^{AC}$ and $\hat{Q}^{-}$. This is a 2D calculation. (E) Mediated energy transfer in a 3-mer along the promoter mode $\hat{Q}^{-}$ is maximized at vibronic resonance. Minimum population on the donor site is plotted as a function of vibrational frequency. The acceptor and intermediate sites are plotted for the time point at which the donor population is minimum.}
	\label{fig:fig5}
\end{figure*}

Fig.~\ref{fig:fig5} presents the main results for the general $\Lambda$-type 3-mer. Panel A shows the relative electronic site energies. The relative site energies and electronic couplings are such that the donor and acceptor excitons are dominantly the same character as the corresponding electronic sites. The resulting donor-acceptor electronic mixing is weak with <5\%  population transfer, while population transfer between the donor-intermediate sites is $\sim$30\%. This is shown in Fig. S2. A straightforward extension to the case of a V system is achieved by changing the relative energy of the intermediate site with respect to the donor and acceptor sites. The corresponding diabatic site potentials along the dominant effective mode $\hat{Q}^{AC}$ are shown in panel B. The vibrational frequency is chosen such that the zero-point levels on the donor and acceptor excitons are separated by a quanta of excitation on the acceptor exciton leading to systematic degeneracies, or vibronic resonances, between the donor-acceptor excitons. Conceptually similar degeneracies and couplings in the context of restricted intramolecular vibrational relaxation (IVR) are known as inter-polyad couplings\cite{Perry2013}. Denoting the excitons with dominantly $A$, $B$ and $C$ character as $\alpha$, $\beta$ and $\gamma$, the approximate analytic forms of the resonantly mixed eigenvectors from the first resonant manifold are --
\begin{eqnarray}
\ket{\pm}&=&\frac{1}{\sqrt{2}}\ket{\alpha}\ket{1^{AC}_{\alpha}}\pm\ket{\gamma}\ket{0^{AC}_{\gamma}}\nonumber\\
\label{eq6}
\end{eqnarray}
The above expression follows from the analysis described in earlier refs.~\cite{Tiwari2018, Tiwari2020, Patra2021}. The vibrational ket $\ket{v^{AC}_{X}}$ corresponds to the vibrational base ket on exciton $X$ along the delocalized effective mode $\hat{Q}^{AC}$ with $v$ quanta of vibrational excitation. The analysis neglects the second order energetic perturbations caused by coupling between neighboring resonant manifolds. Refs.~\cite{Tiwari2020, Patra2021} shows that this assumption holds well for small FC displacements and is sufficient to analytically describe the absorption spectra and population dynamics resulting from exact non-adiabatic eigenvectors obtained from numerical diagonalization of the full Hamiltonian. The lowest eigenvector does not mix with any other exciton and is given by $\ket{\alpha}\ket{0^{AC}_{\alpha}}$. Eqn.~\ref{eq6} implies that excitons $\alpha$ and $\gamma$ are indirectly mixed through a vibronic resonance, even if direct electronic mixing between the donor-acceptor sites is weak. It should be emphasized that this the resonant mixing is only possible due to the presence of both effects - 1. an intermediate site which weakly mediates electronic coupling, 2. vibronic resonance between the donor-acceptor exciton. Any effect alone does not cause such a mixing. We have recently shown\cite{Tiwari2020} that such resonant mixing, when treated without neglecting ground state vibrational excitations, leads to fully delocalized vibronic excitons despite weak initial electronic mixing. Consequently, the vibrational distortion fields around resonantly mixed vibronic excitons are also enhanced. 

It follows that when a coherent superposition of such eigenvectors is excited, $\sim$100\% of the initially excited population is transferred to the acceptor. This is shown in panel B. The population dynamics of the system Hamiltonian is calculated by exciting the donor molecule, and projecting the resulting coherent superposition of the donor-acceptor eigenvectors on the acceptor electronic state. The details of the calculation are briefly described in Section S3. The dramatic enhancement of population transfer essentially reflects the ability of vibronic resonance enhances imperfect delocalization caused by energetic disorder to perfect delocalization\cite{Tiwari2020}. The timescale of this transfer is dictated by the inverse of the $A-C$ domain linear vibronic coupling along $\hat{Q}^{AC}$. Similarly resonant mixing is also expected along $\hat{Q}^{-}$. Intuition gained from symmetric `downhill' and $\Lambda$ Hamiltonians (Section \ref{promoter}) suggests that latter coupling is expected to play only a minor role due to destructive interference between coupling contributions coming from the donor and acceptor FC displacements. This is made evident by a 2D calculation in panel D along both the effective modes, where the amount and timescale of transfer is approximately same as that seen along the dominant effective mode $\hat{Q}^{AC}$.

It should be emphasized that without resonant vibronic mixing, a purely electronic superposition leads to $\sim$30\% population transfer between donor-intermediate sites, with  <5\% transfer to the acceptor. At vibronic resonance the intermediate site switches role from an excitation trap to a mediator for vibronically enhanced population transfer. To emphasized the selectivity of mediated transfer, Panel E shows population on the acceptor and intermediate when the donor population is minimum. The populations are plotted as a function of vibrational frequency. Away from vibronic resonance, almost all the donor population is exchanged with the intermediate site with negligible transfer to the acceptor as expected from $\hat{H}_{elec}$. However, at vibronic resonance, where vibronic couplings become dominant, the population on the intermediate site is minimized, while almost all the donor population is selectively mediated to the acceptor. Interestingly, synergy between vibronic resonance and weak mediated coupling in this mechanism allows for bypassing the intermediate uphill energy transfer all together. This will be recalled in Section \ref{ratchet}.  

In contrast to the $A-C$ energy gap tuning motions which play the dominant role in a general $\Lambda$- system discussed above, for the 'downhill' system, global energy gap tuning motions along  $\hat{Q}^-$ play the major role in mediating energy transfer. This was demonstrated in our earlier study\cite{Patra2021}. The important point is that the switching roles of effective modes is dictated by how electronic Hamiltonian $\hat{H}_{elec}$ rearranges linear vibronic mixing to constructively interfere only along specific effective modes. The above examples of mediated energy transfer in a general $\Lambda$- or V- system suggests that this could be a more general design principle made possible by selectively exploiting vibronic resonances in energetically disordered weakly coupled aggregates with dense FC-active vibrational spectral density, such as photosynthetic proteins. In the following subsections, we further comment on the general aspects of this idea.

\subsubsection{Selectivity of Mediated Energy Transfer}\label{select}

The vibronic eigenvectors in Eqn.~\ref{eq6} suggest that the mediated energy transfer discussed above is independent of whether the intermediate site were optically dark. This is so because the eigenvectors in Eqn.~\ref{eq6} only require the singly-excited electronic state of the donor to be optically allowed for intensity exchange with the first vibronic progression on the acceptor electronic state. Thus the optical properties of the intermediate site do not affect the selectivity of transfer. 

The linear vibronic coupling responsible for mediated energy transfer in a $\Lambda$- or V- system is ultimately arising from the last term in the linear vibronic coupling Hamiltonian $\hat{H}_3^{LV}$ in Eqn.~\ref{eq5}. This vibronic coupling along the dominant $A-C$ tuning mode is independent of the FC displacement on the intermediate site (Eqn.~S28). Thus, even if the  intermediate site has negligible FC displacement along the resonant mode, mediated energy transfer is still possible. 

Instead of optical brightness and FC displacements on the intermediate site, the dominant factor controlling the selectivity of mediated transfer to the acceptor is the relative strengths of vibronic couplings $j_{XY}^{AC}$ between excitons $X-Y$ along the $\hat{Q}^{AC}$ mode -- $j^{AC}_{\alpha\gamma}$ resonant vibronic coupling versus $j^{AC}_{\beta\gamma}$ and $j^{AC}_{\alpha\beta}$ vibronic couplings. The linear vibronic coupling Hamiltonian for the symmetric $\Lambda$ Hamiltonian in Eqns.~S28-S29, shows the vibronic couplings in the respective exciton domains. The amount of mixing is determined by these vibronic couplings and the energy gap between the participating states, and ultimately determines the selectivity of transfer upon donor excitation.

\begin{figure*}[h!]
	\centering
	\includegraphics[width=2.5 in]{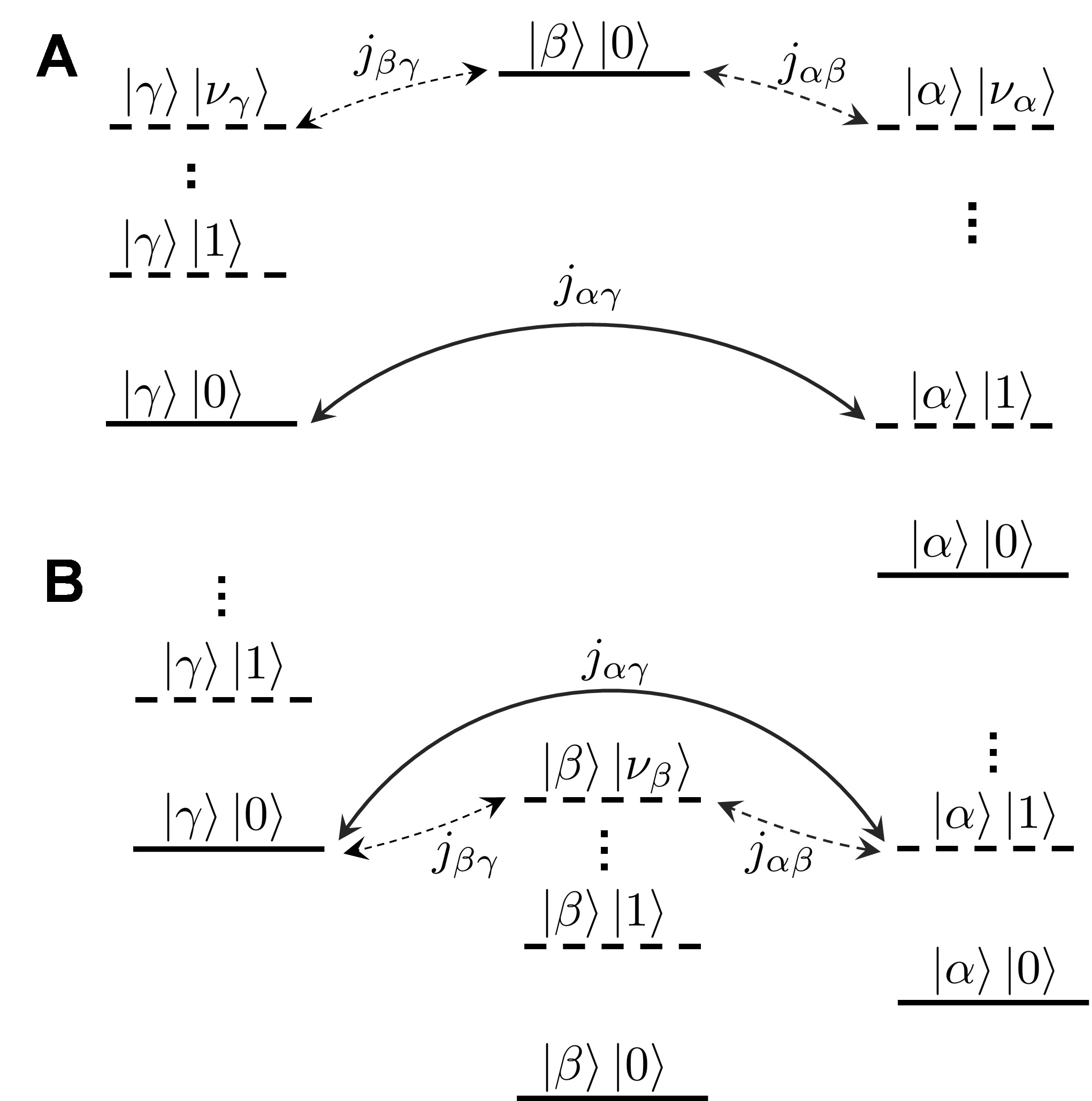}
	\caption{\footnotesize Vibronic mixing and selectivity of population transfer along the dominant promoter mode $\hat{Q}^{AC}$. The strength of vibronic coupling between different exciton domains is schematically denoted as bold and thin dashed arrows to denote strong and weak coupling elements in Eqns.~S28-S29. (A) $\Lambda$ system. Higher resonant manifolds on the donor-acceptor ($\gamma-\alpha$) excitons can mix with the lowest intermediate exciton state. For example, in Fig.~\ref{fig:fig5}B, 3$^{rd}$ progression on the donor can mix, although weakly, with the intermediate exciton. (B) V system. Higher vibrational progression on the intermediate exciton can mix into the lowest resonant donor-acceptor manifold. The vibronic states are denoted by $\ket{X}\ket{v_{X}}$, where $X=\alpha, \beta, \gamma$ excitons and $v_X$ denotes the quanta of vibrational excitation on exciton $X$ along $\hat{Q}^{AC}$. Note that, compared to the text, the superscript $AC$ on vibronic couplings is dropped in the figure for brevity. 
	\label{fig:fig6}}
\end{figure*}

We will utilize the analytic understanding of symmetric $\Lambda$ or V system (Section \ref{promoter}) in order to think about selectivity in the more general $\Lambda$ system. Figure 6 schematically shows the vibronic basis states which dominantly mix together in a general $\Lambda$ versus V system. The relative strengths of vibronic coupling elements for a symmetric $\Lambda$ system (Eqn.~S29) suggests that even in a general system, the ratio of donor-acceptor to donor-intermediate site mixing, that is, $j^{AC}_{\alpha\gamma}/j^{AC}_{\beta\gamma}$ is strongly enhanced due to constructive and destructive interference between intramolecular vibronic coupling elements along the $\hat{Q}^{AC}$ effective mode. Thus, vibronic basis states $\ket{\beta}\ket{v^{AC}_{\beta}}$ and $\ket{\gamma}\ket{v^{AC}_{\gamma}}$ couple very weakly compared to the strongly coupled \textit{resonant} $\gamma-\alpha$ states, determining the selectivity as well as the timescale of population transfer to the acceptor $\alpha$. Similarly, the ratio of donor-acceptor to acceptor-intermediate site mixing is $j^{AC}_{\alpha\gamma}/j^{AC}_{\alpha\beta} \sim \cot(\theta)$, where $\theta$ is the electronic mixing angle for the symmetric $\Lambda$- or V- system defined in Eqn.~S25. Even for moderate-strong electronic mixing angle of 30$^o$, $\gamma-\alpha$ population transfer will be $\sim$3x faster than $\alpha-\beta$ exchange timescale. It is interesting to note how the interplay of $\hat{H}_{elec}$ and the effective mode $\hat{Q}^{AC}$ rearranges vibronic couplings to couple only specific exciton domains. Note that the vibronic couplings $j$ are multiplied by an additional factor coming from the matrix elements of the coordinate vector operator. Thus, difference in vibrational quantum numbers of the vibronically coupled states along with FC displacements determine the total strength of the matrix elements. For weak FC displacements along the delocalized effective modes, this factor is largest when the difference in vibrational quantum numbers is unity.

As depicted in Fig.~\ref{fig:fig6}, a crucial difference can arise between $\Lambda$- and V- systems. For any given set of inter-exciton vibronic couplings, the $\alpha-\beta$ vibronic mixing in the case of V-system imparts a larger $\alpha$ exciton (acceptor) character to $\beta$ (intermediate) exciton. This is because \textit{lower} FC progressions on $\alpha$ (with larger optical intensity) mix with $\beta$. Larger acceptor intensity donated to the intermediate site implies overall reduced selectivity compared to a $\Lambda$ system. Note that in the above analysis it is assumed that exciton energy gap $\Delta_{\alpha,\gamma} \ll \Delta_{\beta,\gamma}$, that is, the $\beta-\gamma$ exciton energy gap is larger by several vibrational quanta along the resonant mode. For instance, for calculations in Fig.~\ref{fig:fig5} these exciton energy gaps differ by $\sim$3x vibrational quanta. Similar selectivity arguments as above can be made in the other limit of exciton energy gaps as well. The interesting case of the two exciton energy gaps being comparable is discussed in the Section \ref{interfere}. \\

\subsubsection{Interference between Vibronic Couplings}\label{interfere}
\begin{figure*}[h!]
	\centering
	\includegraphics[width=5 in]{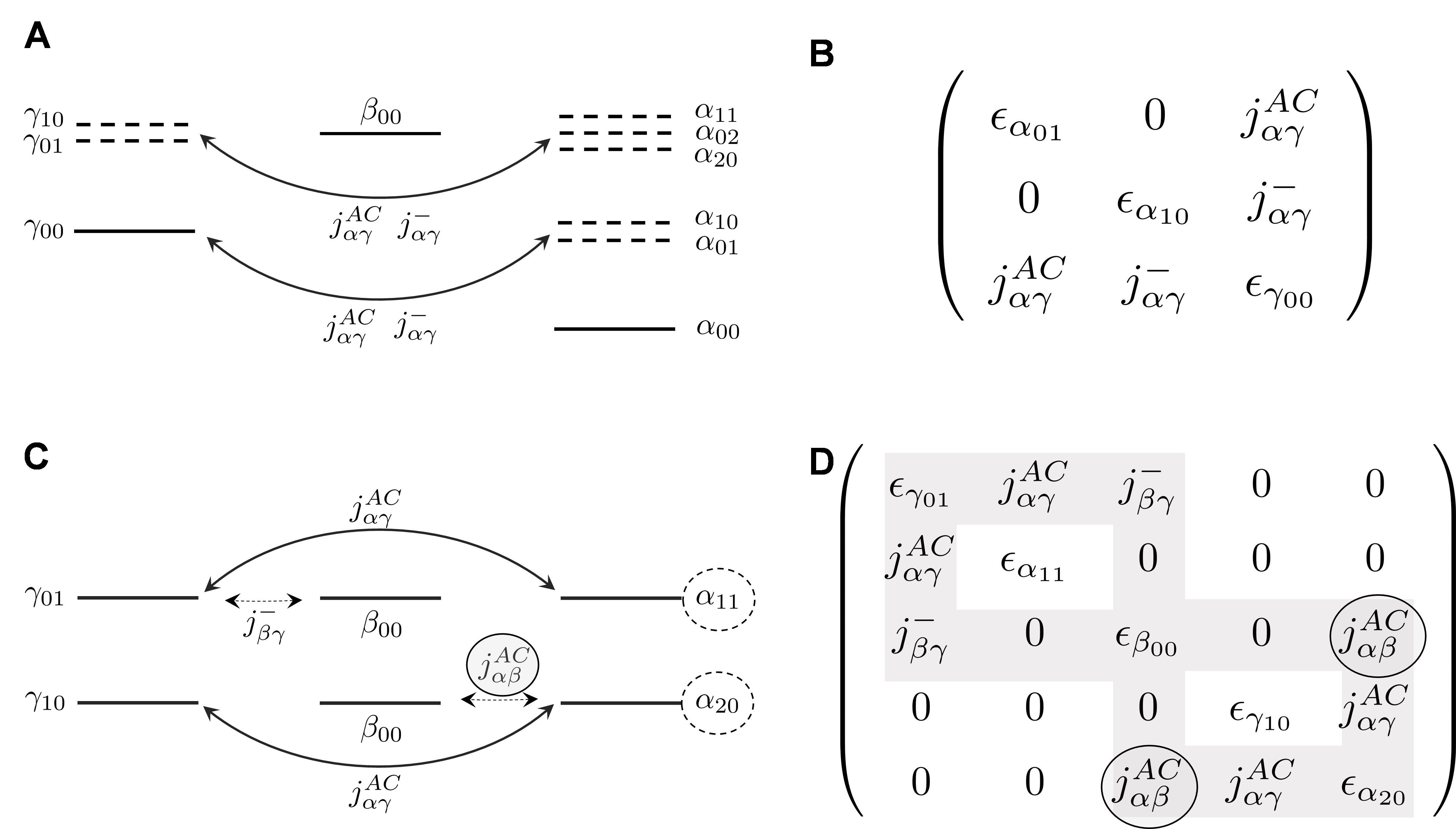}
	\caption{\footnotesize Phase-independent interference between vibronic couplings along different effective modes. (A) Resonant manifolds and vibronic couplings in a $\Lambda$ system. The first resonant manifold with 3 states arises in a general $\Lambda$ system with vibronic resonance. The second manifold, with 5 $\alpha-\gamma$ states resonant with the intermediate state, arises only for the special case of equal exciton energy gaps $\Delta_{\alpha,\gamma} = \Delta_{\beta,\gamma}$. (B) Vibronic couplings in the first resonant manifold derived from Eqns.~S27 and S29. The isoenergetic vibronic basis states are $\alpha_{10}$, $\alpha_{01}$ and $\gamma_{00}$, with corresponding energies $\epsilon_{\alpha_{10}} = \epsilon_{\alpha_{01}} = \epsilon_{\gamma_{00}}$. A rotation of basis about the third domain, decouples one state while increasing the overall vibronic coupling strength between the remaining stats. (C) Dominant vibronic couplings and corresponding isoenergetic basis states in the second resonant manifold. Weaker couplings are denoted by dashed arrows. The circled basis states $\alpha_{11}$ and $\alpha_{20}$ have three-particle character and discussed in Section \ref{npa}. (D) Vibronic couplings corresponding to the second resonant manifold. Stronger coupling elements are shown in bold. The square connecting the matrix elements denotes the 3$\times$3 sub-manifold formally similar to panel B. The vibronic coupling element $j^{AC}_{\alpha\beta}$ is circled to denote that this element is negligible because the matrix element involves 2 quantum of change in the vibrational quantum numbers. 
		\label{fig:fig7}}
\end{figure*}

In Section \ref{met} we showed that for a general $\Lambda$ model, both effective modes $\hat{Q}^{AC}$ and $\hat{Q}^{-}$ can contribute to vibronic mixing with the former being the dominant mode. In the presence of vibronic couplings along both effective modes, the resonant manifold in Eqn.~\ref{eq6} is modified by the addition of a third vibronic basis state along $\hat{Q}^{-}$. The resulting resonant basis states $\ket{\alpha}\ket{1^{AC}_{\alpha}}\ket{0^{-}_{\alpha}}$, $\ket{\alpha}\ket{0^{AC}_{\alpha}}\ket{1^{-}_{\alpha}}$ and $\ket{\gamma}\ket{0^{AC}_{\gamma}}\ket{0^{-}_{\gamma}}$ will be denoted as $\alpha_{10}$, $\alpha_{01}$ and $\gamma_{00}$. Figure \ref{fig:fig7}A schematically shows the resonant manifolds possible in a general $\Lambda$ system with exciton energy gaps $\Delta_{\alpha,\gamma} = \Delta_{\beta,\gamma}$. Analyzing the corresponding vibronic couplings in a symmetric $\Lambda$ system (Eqns.~S27 and S29) can again be useful. These are shown for the first resonant manifold in Fig.~\ref{fig:fig7}B. The interference between vibronic couplings is seen by a basis set rotation about the 3$^{rd}$ electronic domain ($\gamma$) with a rotation angle $\eta = \arctan(j^-_{\alpha\gamma}/j^{AC}_{\alpha-\gamma})$.  The relative strengths of vibronic couplings $j^{AC}_{\alpha\gamma}$ and $j^{-}_{\alpha\gamma}$ in a symmetric $\Lambda$ system imply that mixing angle $\eta$ is expected to be small even in a general system. Such a transformation decouples one state, while coupling the state with dominant $\alpha_{10}$ character to $\gamma_{00}$ with \textit{larger} vibronic coupling given by $\sqrt{(j^{AC})^2 + (j^{-})^2}$. This is formally similar to the transformation that decouples\cite{Tiwari2018, Tiwari2020} correlated vibrations from the vibronic coupling Hamiltonian. Interestingly, the interference between vibronic couplings described here is not phase dependent and always leads to larger overall couplings. This is distinctly different, for example, from the case of null excitons\cite{Hariharan2021} where charge-transfer and Coulomb couplings can interfere destructively. 

The phase-independent interference between vibronic couplings manifests in more general contexts. In the context of multiple intramolecular vibrational modes in an excitonic dimer, ref.\cite{Tiwari2018} has shown that depending on the width of vibronic resonance, vibronic coupling along multiple near-resonant (not exactly resonant) tuning modes can interfere constructively to result in larger overall vibronic coupling. For example, see Fig.~3 and Fig.~5 of ref.\cite{Tiwari2018}. Extending this reasoning to the case of $\Lambda$ 3-mer, in the presence of an additional near-resonant intramolecular mode, the constructive interference in panel A will be further enhanced with overall coupling $\sqrt{\sum_{n=1}^{2}\big[(j^{AC}_n)^2 + (j^{-}_n)^2\big]}$. This can be easily shown using transformations similar to panel A (see Eqn.~S6 of Sahu et al.\cite{Tiwari2020} for a related transformation). Enhancing the vibronic couplings arising from individual effective modes and near-resonant modes is one of the ways mutual interference of vibronic couplings leads to faster population transfer rates. The interference of vibronic couplings along multiple modes implies that vibrational motions along orthogonal effective modes can simultaneously drive the vibronic probability density evolution between electronic states. This is similar to the case of indirect coupling between independent vibrational modes discussed by Makri and co-workers\cite{MakriARPC2022}. This will be recalled in Section \ref{5mer}.

The higher resonant manifold between $\gamma-\alpha$ excitons in Fig.\ref{fig:fig7}A  becomes resonant with $\beta_{00}$ when exciton energy gap $\Delta_{\alpha,\gamma} = \Delta_{\beta,\gamma}$. Now the intermediate exciton can resonantly mix with the donor and acceptor excitons, and one may expect less selective donor-acceptor energy transfer. However, interference between vibronic couplings can still channel energy transfer selectively to the acceptor. Fig.~\ref{fig:fig7}C shows the isoenergetic basis states in the second resonant manifold which are coupled by dominant vibronic couplings $j^{AC}_{XY}$ and $j^{-}_{XY}$. Taking clues from the symmetric $\Lambda$ model, for equal intramolecular FC displacements $j^{-}_{\alpha\gamma}$, $j^{-}_{\alpha\beta}$ and $j^{AC}_{\beta\gamma}$ are exactly zero. The resulting dominant couplings are shown in Fig.~\ref{fig:fig7}C. The highlighted coupling element $j^{AC}_{\alpha\beta}$ involves change in vibrational quanta of 2 and will be negligible even for moderately large intramolecular FC displacements considered here.  From the discussion in Section \ref{select}, $j^{AC}_{\alpha\gamma}/j^{AC}_{\alpha\beta} \sim \cot(\theta)$, such that $\alpha-\beta$ coupling is weak even for moderate electronic mixing. This ratio is further strongly enhanced due to a change of 2 vibrational quanta in $j^{AC}_{\alpha\beta}$. Eqns.~S27 and S29 show that $j^{-}_{\beta\gamma}$ is weaker than the dominant $j^{AC}_{\alpha\beta}$ coupling by a factor of $\cos(\theta)/\sqrt{3}$. Panel D shows the resulting coupled manifold Hamiltonian, where 3$\times$3 sub-domains similar to panel A are highlighted. Using transformations similar to those discussed in panel B, it can be shown that, as before, vibronic couplings along effective modes interfere as a square. This is also easily seen if the highlighted $j^{AC}_{\alpha\beta}$ is ignored. This underscores the earlier point that vibronic couplings along effective modes are partitioned such that only specific electronic domains are strongly coupled. The interference between vibronic couplings ensures mixing caused by weaker couplings is further suppressed. Thus, both effects together allow for selective energy transfer to the acceptor, despite a vibronic resonance with the intermediate exciton. It is interesting to note that such interference effects can also couple vibronic resonances between different pairs of excitons arising from multiple vibrational modes. This will be a subject of our forthcoming publication. A related example is analyzed in our recent work\cite{Patra2021} on a downhill 3-mer, where a \textit{common} vibrational frequency couples multiple vibronic resonances (see Section S12 of Patra et al.\cite{Patra2021}).

\subsubsection{Bypassing Uphill Energy Transfer}\label{ratchet}

Donor-acceptor energy transfer in a $\Lambda$ system involves an uphill energy transfer step, and therefore may be inefficient in the presence of weak to intermediate electronic mixing. An uphill step could be efficiently overcome via coherent electronic coupling large enough to strongly mix the donor-acceptor sites via the intermediate site. However, an undesirable consequence of a mechanism relying on large mediated electronic mixing is the undesirable substantial population exchange with the intermediate site. As a promising alternative, Sections \ref{select} and \ref{interfere} show that partitioning of linear vibronic coupling along specific effective modes and interference between vibronic couplings can be leveraged to achieve selective donor-acceptor energy transfer that can be substantially enhanced at vibronic resonance. The uphill energy transfer step to the intermediate site is \textit{bypassed} in this mechanism, such that the role of intermediate site is minimal and amounts to mediating weak electronic mixing between the donor-acceptor sites.

The mechanism for mediated energy transfer proposed above could serve as an interesting design principle for efficiently mediating energy transfer by \textit{skipping} uphill energy transfer steps all together. Energetically disordered aggregates with dense vibrational spectral density may be promising candidates to explore such mechanisms.  Mediated vibronic coupling may be possible in singlet exciton fission\cite{Michl2013} where charge-transfer (CT) states can be strongly coupled with (mutually weakly coupled) locally excited (LE) and correlated triplet (TT) states. However, owing to substantially higher lying CT states, the overall electronic mixing between LE and TT states may be too weak to promote efficient ultrafast formation of TT state. \\ 

Note that the mechanism of bypassing uphill energy transfer discussed here is distinctly different from leveraging vibronic resonances as a quantum ratchet\cite{Fleming2020_2} to promote uphill energy transfer. For efficient uphill energy transfer to occur, the initial optical excitation probability of vibronically mixed eigenvectors in Eqn.~\ref{eq6} becomes key. For example, $\bra{G}\hat{\mu}\ket{\alpha}\bra{0^{AC}_G}\ket{1^{AC}_{\alpha}}$ and $\bra{G}\hat{\mu}\ket{\gamma}\bra{0^{AC}_G}\ket{0^{AC}_{\gamma}}$ become comparable for large intramolecular FC displacements. In this regime, the energy transfer can be mediated to and from the acceptor with equal probability thus enabling uphill energy transfer via a vibronic resonance. Peters et al. have also discussed\cite{Peters2017} that in such cases energy transfer through a nested funnel becomes less directional. Note that following the same argument, a combination of higher temperatures and low frequency vibrations such that occupation probability of $v=$1 levels on the ground state is large can also allow efficient uphill energy transfer. Ultimately the timescale of vibronic decoherence\cite{Tiwari2018} caused by the bath becomes crucial for mediated or uphill energy transfer to be realized.

\subsection{Mediated Energy Transfer in Larger Aggregates : $\Lambda$ 5-mer}\label{5mer}

\begin{figure*}[h!]
	\centering
	\includegraphics[width=5 in]{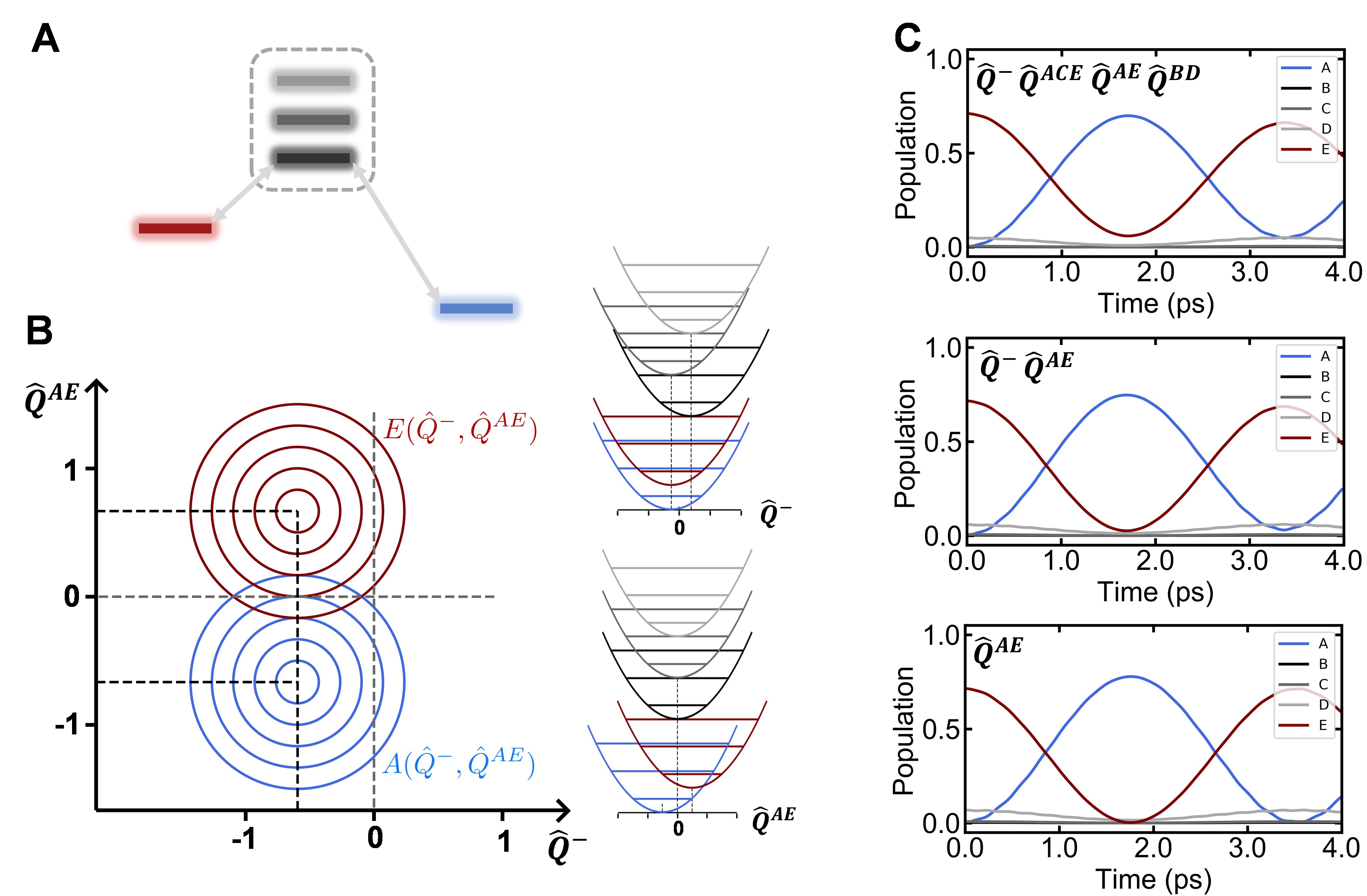}
	\caption{\footnotesize (A) Relative electronic site energies of $\Lambda$ 5-mer system. The intermediate sites are highlighted by black dashed box. The intermediate sites are mutually strongly coupled, while the donor and acceptor sites are not directly coupled. They are both weakly coupled to one of the intermediate sites as shown by connecting arrows. The parameters are listed in Fig.~S2. Each electronic site has one identical intramolecular FC active vibration of frequency 125 cm$^{-1}$ and FC displacement $d=\sqrt(0.5)$, such that each electronic domain in the 5-mer Hamiltonian is 5D. (B) 2D diabatic site potentials corresponding to the singly-excited acceptor and donor electronic states $A$ and $E$ respectively. The potentials are plotted along the dominant effective modes $\hat{Q}^-$ and $\hat{Q}^{AE}$, and denoted as $X(\hat{Q}^{AE}, \hat{Q}^-$), where $X=A,E$ denotes the singly-excited donor and acceptor electronic states. The diabatic potential surfaces are only coupled indirectly through Coulomb coupling with the intermediate sites. Intermediate sites are not shown for clarity. The FC displacements along the effective modes in the diabatic sites basis are listed in Table S2. 1D diabatic site potentials for each of the 5 singly-excited electronic states along the two dominant effective modes are also shown. The FC displacements and relative energies are shown to scale. The vertical dotted lines mark the FC displacements on the coordinate axis. All intermediate potentials are shown in dark gray. (C) Population dynamics of the $\Lambda$ 5-mer Hamiltonian after exciting a superposition of vibronically mixed donor and acceptor states upto the 1$^{st}$ resonant manifold in panels B,C. (Top) 4D calculation along all effective modes except the global correlation vector $\hat{Q}^{+}$, with vibrational basis $\ket{v^-}\ket{v^{AE}}\ket{v^{ACE}}\ket{v^{BD}}$. (Middle) 2D calculation along the promoter modes $\hat{Q}^{-}$ and $\hat{Q}^{AE}$ with vibrational basis $\ket{v^-}\ket{v^{AE}}$. (Bottom) 1D calculation along the dominant mode $\hat{Q}^{AE}$ with vibrational basis $\ket{v^{AE}}$.} 
	\label{fig:fig8}
\end{figure*}

Sections \ref{normalmodes} and \ref{iterative} discuss the structure of effective modes for larger aggregates. Based on the schematic shown in Fig.~\ref{fig:fig2}A, for any given set $j$ of intramolecular FC modes, the 5-mer effective modes  are $\hat{Q}^{+}$, $\hat{Q}^{-}$, $\hat{Q}^{ACE}$, $\hat{Q}^{AE}$ and $\hat{Q}^{BD}$. The singly-excited electronic states are denoted as $A-E$, where $A$ implies site $A$ excited while all other sites in their ground electronic state. The derivation of 5-mer effective modes is detailed in Section S1.2, and the diabatic site FC displacements along these modes are listed in Table S2. The 5-mer electronic Hamiltonian is shown in Fig.~S2, and chosen such that the donor and acceptor are only indirectly coupled through weak electronic coupling mediated by the intermediate sites. Fig. \ref{fig:fig8}A shows the relative electronic site energies on scale. The intermediate sites are mutually strongly coupled, and the donor and acceptor are coupled to one of the intermediate sites. The electronic site energies and couplings are such that upon donor excitation, <5\% excitation is transferred to the acceptor, while $\sim$40\% population is exchanged with the intermediate sites. This is shown in Fig.~S2. Each molecule has one identical FC active vibrational mode, such that the Hamiltonian for each electronic state of the system, including the ground electronic state, is 5D. As in Section \ref{met}, the frequency of the vibrational mode is chosen so as to have a vibronic resonance between the donor-acceptor excitons. As we will see numerically, the dominant effective modes in this problem is only the $A-E$ energy gap tuning mode $\hat{Q}^{AE}$ with a minor role for global tuning mode $\hat{Q}^-$, such that the overall donor-acceptor vibronic mixing can be very well approximated with a 2D problem along the promoter modes. 

Fig.~\ref{fig:fig8}B plots the 2D diabatic site potentials along the dominant effective modes, for donor and acceptor singly-excited electronic states $E(\hat{Q}^{AE},\hat{Q}^{-})$ and $A(\hat{Q}^{AE},\hat{Q}^{-})$, respectively. The relative FC displacements along each effective mode of the 5-mer are listed in Table S2, are shown on scale in the figure. Recalling the discussion of interference of vibronic couplings in Section \ref{interfere}, linear vibronic couplings along all effective modes influence vibronic probability density evolution on any given excited electronic state, although on timescales depending on the individual strengths of vibronic couplings. In the case of $\Lambda$ 5-mer, vibronic couplings along the dominant effective modes, $\hat{Q}^{AE}$ and $\hat{Q}^{-}$ will interfere (Section \ref{interfere}) to simultaneously dictate the short-time evolution of the vibronic probability density, which could to a good approximation be visualized using the 2D diabatic potentials (Fig.~\ref{fig:fig8}B) made possible through the effective mode transformation.  The 1D diabatic potentials individually along the promoter modes $\hat{Q}^{AE}$ and $\hat{Q}^{-}$ are also shown. The relative FC displacements between the states determines the strength of vibronic coupling. It can be seen that the relative displacement between the donor and acceptor electronic states is largest along $\hat{Q}^{AE}$. Recalling the discussion in Section \ref{normalmodes}, the effective mode transformation partitions the vibronic couplings such that smaller FC displacements are expected along more delocalized global tuning mode $\hat{Q}^-$ compared to pairwise tuning mode $\hat{Q}^{AE}$ leading to stronger vibronic coupling. As in the case of 3-mer, even though the donor-acceptor electronic potentials are not relatively tuned along $\hat{Q}^-$, it is the tuning relative to the intermediate sites that is responsible for vibronic coupling along the global mode. As we have seen for the case of 3-mer (Section \ref{promoter}), given the vibronic couplings along each effective mode, it is ultimately $\hat{H}_{elec}$ which \textit{determines} the dominant promoter mode. 

A dominant role for the effective mode which directly tunes the $A-E$ electronic energy gap compared to the global tuning mode is not surprising if one recalls the physical intuition gained from the 3-mer $\Lambda$ model. Further, interference effects between vibronic couplings (Section \ref{interfere}) suggests a suppression of weaker couplings in the dynamics. With these expectation, Fig.~\ref{fig:fig8}C compares the population dynamics of the $\Lambda$ 5-mer for the full 4D calculation versus calculations with reduced dimensionality along the dominant promoter modes. The calculations are described in Section S3. The 4D calculation, shown in the top panel, is along the effective modes $\hat{Q}^{-}$, $\hat{Q}^{AE}$, $\hat{Q}^{ACE}$ and $\hat{Q}^{BD}$ and considers vibronic coupling along \textit{all} effective modes. Note that the global correlation mode $\hat{Q}^{+}$ is separable from the electronic Hamiltonian and plays no role in vibronic mixing. Similar to 3-mer, the $\sim$100\% donor-acceptor population exchange becomes possiblethrough the combined effect of weak electronic coupling and vibronically resonant donor-acceptor excitons. In line with the above expectations, the timescale of population transfer, which relates to the total vibronic coupling strength, is approximately equal even for reduced calculations along $\hat{Q}^{-}$, $\hat{Q}^{AE}$ (middle) and along $\hat{Q}^{AE}$ (bottom) alone. This suggests that $\hat{Q}^{AE}$ indeed is the dominant promoter mode and influences dynamics on the fastest timescales. 1D calculations along individual effective modes are shown in Figure S3. The above analysis of the general 5-mer $\Lambda$ system using the effective mode transformation allows for identifying the dominant promoter mode, and therefore reducing the system dimensionality without severely approximating vibronic coupling and associated timescales. The generality of mediated energy transfer in the presence of multiple intermediate sites is further confirmed by Fig.~S4, where again vibronic resonance selectively enhances direct donor-acceptor transfer while bypassing intermediate uphill energy transfer steps.

\subsubsection{Role of n-particle basis states}\label{npa}

As discussed in Section \ref{theory}, resonant vibronic mixing requires explicit quantum treatment of vibrations in the system Hamiltonian. In case of extended aggregates, this enforces truncation of multi-particle basis sets for a feasible computation time. Often truncation up to 2-particle basis states is adequate to describe linear spectral lineshapes\cite{Hestand2018,Painelli2019}. However, from a quantum dynamical perspective, resonant vibronic mixing enhances multi-particle basis states contributions through intensity borrowing. Description of  fundamental properties of vibronic excitons such as delocalization, vibrational distortion radius, and energy transfer rates crucially depends\cite{Tiwari2020} on multi-particle states.   

Although multiple exciton energy gaps are present in the $\Lambda$ system treated in Sections \ref{promoter} and \ref{5mer}, only one vibronic resonance between the donor-acceptor excitons exists. This may make $\Lambda$ Hamiltonians seem similar to a dimer where 2-particle basis states can completely describe the system. However, this is not the case. Consider the $\Lambda$ system depicted in Fig.~\ref{fig:fig7}, where exciton energy gaps are comparable to vibrational quanta. Recalling the discussion in Section \ref{interfere}, the dominant vibronic coupling $j_{\alpha\gamma}^{AC}$ dictates the selectivity of mediated energy transfer. As shown in Fig.~\ref{fig:fig7}C, vibronic excitons $\alpha_{11}$ and $\alpha_{20}$ couple with the donor exciton $\gamma$ through two mixing channels. Using transformations from intramolecular $\ket{v_{A}}\ket{v_{B}}\ket{v_{C}}$ to delocalized  $\ket{v^{+}}\ket{v^{-}}\ket{v^{AC}}$  vibrational basis, it can be shown that vibrational basis states $\ket{0^{+}}\ket{1^{-}}\ket{1^{AC}}$ and $\ket{0^{+}}\ket{0^{-}}\ket{2^{AC}}$ have substantial 3-particle character. For example, following the analysis similar to Section S1 of ref. \cite{Tiwari2018},
\begin{eqnarray}
\ket{\alpha}\ket{0^+0^-2^{AC}}&=&\frac{1}{2}\ket{\alpha}\big[\ket{2_A0_B0_C}
+\ket{\alpha}\ket{0_A0_B2_C}\big]\nonumber\\
&+&\frac{1}{\sqrt{2}}\ket{\alpha}\ket{1_A0_B1_C}\nonumber\\
%\scalemath{0.8}{
	\ket{\alpha}\ket{0^+1^-1^{AC}}&=&\frac{1}{\sqrt{6}}\ket{\alpha}\big[\ket{2_A0_B0_C}-\ket{0_A0_B2_C}\big]\nonumber\\
	&+&	\frac{1}{\sqrt{3}}\ket{\alpha}\big[\ket{0_A1_B1_C}-\ket{1_A1_B0_C}\big],\nonumber\\
	\nonumber
\end{eqnarray}
where the 2$^{nd}$ term in both equalities indicates a substantial 3-particle character with vibrational excitations on two electronically unexcited states. Thus, it is fairly counter-intuitive that even a 3-mer with only one vibronic resonance requires 3-particle basis states to correctly describe interference between vibronic couplings and  resulting selectivity of mediated transfer. Analogous reasoning for the 5-mer in Fig.~\ref{fig:fig8} also suggests a vital role of multi-particle basis states, the essential idea stemming form the fact that ground state vibrational excitations on all coupled molecules contribute to any given singly-excited electronic states. \\

\section{Conclusions}\label{conclusions}

We have extended the effective mode approach to highlight the physically intuitive and iterative nature of effective normal modes for successively larger aggregates. By applying the effective mode transformation, a multidimensional energy transfer Hamiltonian is written as exactly as a sum of 1D Hamiltonians along each of the effective modes. The effective mode transformation partitions the linear vibronic coupling arising from intramolecular FC vibrational modes along specific effective modes. 

Through analytic transformations we have illustrated how the interplay of electronic Hamiltonian and the effective mode transformation selects the dominant promoter mode. Our results suggest an interesting possibility of engineering electronic Hamiltonians to drive vibronic probability density along specific directions in the multidimensional excited state electronic potentials. It should be emphasized that these physical insights are made possible by the ability to analyze vibronic mixing contributions along specific effective modes.

Using the effective mode formalism, we have generalized the concept of mediated energy transfer to $\Lambda$ or V type molecular aggregates. We have shown that synergy between weak mediated coupling and vibronic resonance allows for skipping any intermediate uphill energy transfer steps and enhancing vibronic mixing to near-perfect exciton delocalization. Interestingly, phase-independent interference between vibronic couplings along different effective modes enhances the dominant vibronic coupling along the promoter mode while suppressing weaker contributions from other effective modes. Such interference effects determine the selectivity of mediated energy transfer. 

The approach presented here is a promising route for four-wavemixing spectroscopic simulations of excitonically coupled aggregates where rich spectroscopic signatures and interference effects\cite{Policht2022,Palecek2017} arising from dominant vibronic couplings are expected. This will be subject of future applications of this approach. Typically one-particle approximation to resonant vibronic coupling, although a severe approximation\cite{Tiwari2020}, becomes unavoidable for simulating such systems with practical computation times. Further, the $\Lambda$- and V- type systems analyzed here can be fairly common in energetically disordered multi-chromophoric aggregates such as photosynthetic proteins and organic thin films where multiple near-resonant vibrational modes and exciton energy gaps are expected. Our analysis suggests that inclusion of multi-particle basis states (beyond two-particle approximation) becomes vital for calculations describing a functional role of vibronic mixing and coherences in enhancing energy and charge delocalization.

\section*{Acknowledgments}
SP acknowledge research fellowship from the Indian Institute of Science. VT acknowledges the Infosys Young Investigator Fellowship supported by the Infosys Foundation, Bangalore. This project is supported by Science and Engineering Research Board, India under grant sanction numbers CRG/2019/003691 and IPA/2020/000033, and Department of Atomic Energy, India under grant sanction number 58/20/31/2019-BRNS. 
 
\section*{Data Availability}
The data that support the findings of this study are available from the corresponding author upon reasonable request 

\bibliographystyle{unsrt}
\bibliography{EMA2Refs}

\end{document}